\pdfoutput=1
\documentclass[10pt,aps,pre,amsmath,amsfonts,amssymb,floatfix,longbibliography,superscriptaddress,notitlepage, twocolumn]{revtex4-1}
\usepackage{mathrsfs} \usepackage{graphicx,color} \usepackage{bbm} \usepackage{multirow}
\usepackage[normalem]{ulem}
\usepackage{lineno}
\usepackage{empheq}
\usepackage{float}

  \def\erf{\text{erf}}

\newcommand{\blue}[1]{\textcolor{black}{#1}}

\newcommand{\magenta}[1]{\textcolor{black}{#1}}

\def\to{\rightarrow}

\newcommand{\beq}{\begin{equation}} \newcommand{\eeq}{\end{equation}}

\renewcommand{\phi}{\varphi}

\renewcommand{\vec}[1]{\boldsymbol{\mathrm{#1}}}
\def\vr{{\boldsymbol r}}

\newcommand\ESI{ESI\dag}

\begin{document}

\title{Machine learning glass caging order parameters with an artificial nested neural network}

\author{Kaihua Zhang}
\altaffiliation{Contributed equally to this work}
\affiliation{School of Chemistry, Beihang University, Beijing 100191, China}

\author{Xinyang Li}
\altaffiliation{Contributed equally to this work}
\affiliation{CAS Key Laboratory of Theoretical Physics, Institute of Theoretical Physics, Chinese Academy of Sciences, Beijing 100190, China}
\affiliation{School of Physical Sciences, University of Chinese Academy of Sciences, Beijing 100049, China}

\author{Yuliang Jin}
\email{yuliangjin@mail.itp.ac.cn}
\affiliation{CAS Key Laboratory of Theoretical Physics, Institute of Theoretical Physics, Chinese Academy of Sciences, Beijing 100190, China}
\affiliation{School of Physical Sciences, University of Chinese Academy of Sciences, Beijing 100049, China}

\author{Ying Jiang}
\email{yjiang@buaa.edu.cn}
\affiliation{School of Chemistry, Beihang University, Beijing 100191, China}
\affiliation{Center of Soft Matter Physics and Its Applications, Beihang University, Beijing 100191, China}

\begin{abstract}
Around a glass transition, the dynamics of a supercooled liquid dramatically slow down, exhibited by caging of particles, while the structural changes remain subtle.
In alternative to recent machine learning studies searching for structural predictors of glassy dynamics, here we propose  to learn directly particle caging features defined purely according to dynamics.
We focus on three transitions in a simulated hard sphere glass model, the melting of ultra-stable glasses, the Gardner transition and the liquid to ordinary glass transition. Implementing the machine learning algorithm   based on a two-level nested neural network, we attain not only proper caging order parameters for all three transitions, but also a phase classification for input samples. A finite-size scaling analysis of the phase classification results identifies 
 the order of melting (first) and Gardner (second) transitions. A false positive  is avoided, as the liquid to glass transition is indicated as a crossover, rather than a phase transition with a well-defined transition point. This study paves the way to a generic approach for learning dynamical features in glassy systems, with a minimum requirement of system-specific knowledge.
\end{abstract}

\maketitle

\section{Introduction}

The essential idea of Landau theory 
is to define suitable order parameters for phase transitions;
 e.g., 
 a liquid-to-crystal transition can be  characterized by order parameters that quantify the corresponding crystal structure. 
However, the problem is more complicated
{ for}
transitions in disordered systems, such as 
glasses.
For instance, around the liquid to glass transition, 
the viscosity increases by more than ten orders of magnitude over a small range of temperatures~\cite{angell1995formation}, while both phases seem to remain ``disordered".

Great efforts have been devoted to finding a proper glass order parameter based on static  configurations of particles
~\cite{royall2015role, tanaka2019revealing, steinhardt1983bond,xia2015structural, marin2020tetrahedrality,leocmach2012roles,hirata2011direct, hu2015five, taffs2016role,tong2018revealing,tong2019structural,miracle2004structural}. 
Besides the physics approach, recently researchers have attempted 
to address the conundrum by taking advantage of the power of machine learning (ML) techniques in extracting hidden information from  structural data.
A parameter called ``softness" was learned to predict structural flow defects and plasticity in amorphous solids~\cite{cubuk2015identifying,schoenholz2016structural, cubuk2017structure, schoenholz2017relationship}. Graph neural networks (GNN) were used to predict long-time dynamics  
from  the graph structure of  initial particle positions~\cite{bapst2020unveiling}. Inspired by the GNN method, new structural descriptors were designed by recursively incorporating averaged features from neighbour shells~\cite{PhysRevLett.127.088007}. Unsupervised ML methods were also developed to identify different clusters of particles based on their static correlations~\cite{paret2020assessing, boattini2020autonomously}, or network representations that encode  interatomic potentials~\cite{ronhovde2011detecting, ronhovde2012detection}. 

In parallel to the static approach of structural order parameters, alternatively, one can consider glass {\it caging order parameters} defined purely based on dynamics~\cite{parisi2010mean, parisi2020theory}.
\blue{Compared to structural order parameters, which are often system-specific, caging order parameters seem more generic.}
 For example, 
 in spin glasses, where dynamics are usually decoupled from structures, it is standard to define and analyze  overlapping order parameters~\cite{mezard1987spin}, which play a similar role as the caging order parameters in structural glasses~\cite{parisi2020theory}. 
The caging/overlapping order parameters are at the core of  mean-field theories of glass transitions, including the mode-coupling theory~\cite{gotze2008complex, reichman2005mode}, the replica theory~\cite{mezard1987spin, parisi2010mean, parisi2020theory}, and the random first-order phase transition theory~\cite{kirkpatrick1987connections, kirkpatrick1989scaling}. \blue{To this day, it remains unexplored whether caging order parameters can be  identified by ML.}

Inspired by mean-field glass theories, in this 
{ article}
we design a two-level nested neural network (NNN) to learn directly  caging features in glasses. 
Such a dynamics-oriented ML strategy does not rely on any structures.
\magenta{Upon training the model to achieve a phase classification~\cite{ Carrasquilla2017Machine}  by the second-level network, the front-connected  networks at the first level autonomously extract the inherent 
{ features}
beneath the replicated input configurations, physically interpreted as caging order parameters.}
\blue{
Applying a proper finite-size scaling analysis to the machine provided classifications can further identify the order of phase transitions, a method that, to our knowledge, was previously unavailable. We demonstrate the above abilities by applying our method to melting and Gardner transitions in ultra-stable hard sphere (HS) glasses. 
In addition, we show that dynamical crossovers (e.g., the liquid to glass transition) can be distinguished from true phase transitions within this framework. }

\section{System and transitions}
We consider a polydisperse HS glass model (see Electronic Supplementary Information (ESI)\dag~ {Sec.~S1} for details)
whose mode-coupling theory (MCT) transition {temperature} is at $\hat{T}_{\rm MCT} \approx 0.044$~\cite{berthier2016growing} (or {volume fraction} $\varphi_{\rm MCT} \approx 0.594$, in this study, we use reduced temperatures).
Two kinds of glasses, ordinary and ultra-stable glasses, are numerically created.  
The ordinary glasses are prepared by a moderate compression rate $\Gamma = 10^{-3}$ using molecular dynamics (MD) simulations,  corresponding to a glass transition temperature  $\hat{T}_{\rm g} \approx 0.047$ (or $\varphi_{\rm g} \approx 0.58$).
The ultra-stable glasses  with a glass transition temperature   $\hat{T}_{\rm g} \approx 0.033$ (or $\varphi_{\rm g} \approx 0.63$) are generated by an efficient swap Monte Carlo algorithm~\cite{berthier2016equilibrium}(see \ESI~{Sec.~S2}). 
We prepare about 2000 uncorrelated equilibrium configurations at $\hat{T}_{\rm g}$, which are referred to as $\it samples$ in the following.  Except for the preparation of  ultra-stable initial states, all other simulations are performed using pure MD without swap.

We study three transitions:
 (i) The melting of ultra-stable HS glasses at the melting temperature  $\hat{T}_{\rm m} (\hat{T}_{\rm g} )>\hat{T}_{\rm g}$, by decompression with $\Gamma = - 10^{-4}$ from the deep equilibrium states at $\hat{T}_{\rm g} \approx 0.033$. Recent studies suggested that the discontinues  melting of ultra-stable glasses is a vestige of a hidden first-order phase transition taking place in two coupled replicas of the system~\cite{jack2016melting, berthier2015evidence}. 
(ii) The Gardner transition in ultra-stable HS glasses at $\hat{T}_{\rm G} (\hat{T}_{\rm g})<\hat{T}_{\rm g}$, by compression with $\Gamma = 10^{-4}$ from $\hat{T}_{\rm g} \approx 0.033$. The Gardner transition is a second-order phase transition separating the {\it stable glass} (at $\hat{T}> \hat{T}_{\rm G}$)  and the {\it marginally stable  glass} (at $\hat{T}< \hat{T}_{\rm G}$)  phases~\cite{charbonneau2014fractal, parisi2020theory,berthier2016growing,charbonneau2017nontrivial,li2021determining}. (iii) The liquid to ordinary glass transition  around $\hat{T}_{\rm g} \approx 0.047$, which is believed to be a dynamical crossover rather than a true phase transition.

 \begin{figure}[ht]
\centerline{\includegraphics[width=1.\columnwidth]{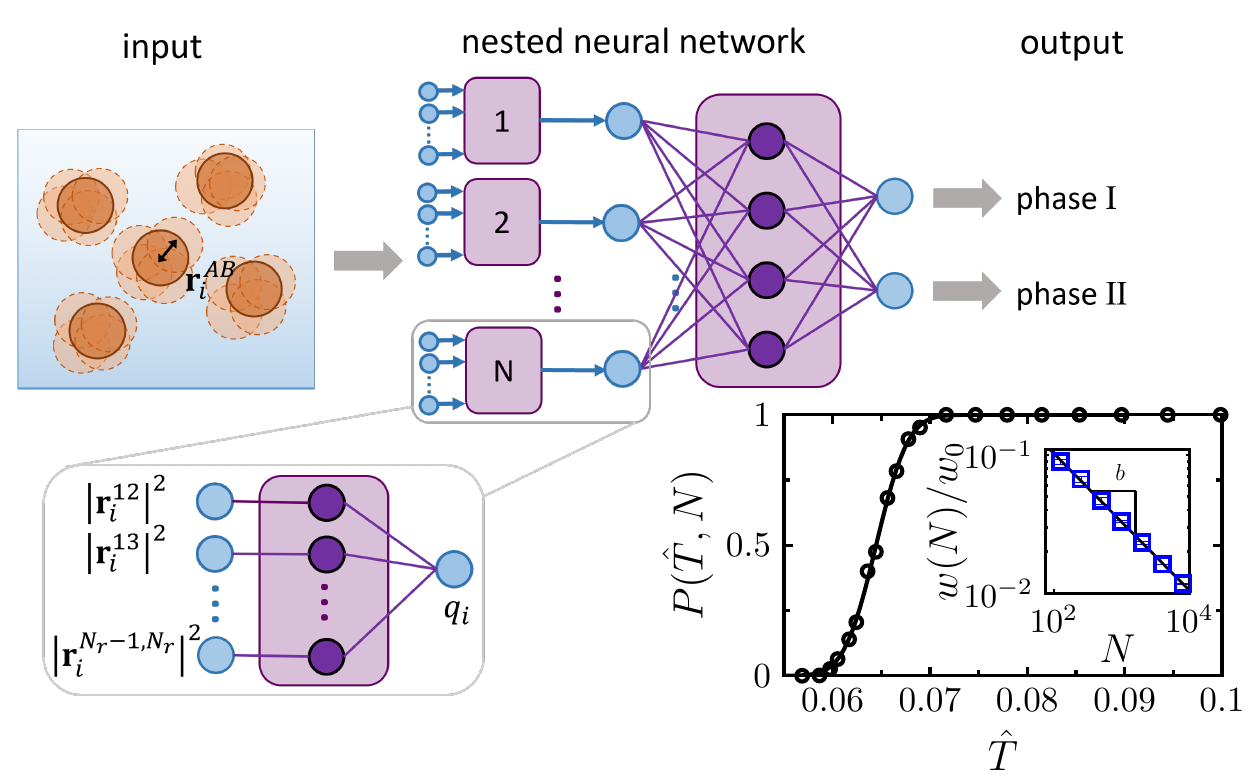}} 
\caption{{Schematic diagram of the machine learning method.}
An example of the output $P(\hat{T}, N)$ is shown for the melting transition ($N=125$), where the line represents the fitting to Eq.~(\ref{eq:ML_fitting}). (inset) Fitting the transition width $w(N)$ to Eq.~(\ref{eq:ML_w_scaling}), giving $b=0.496(5)$.}
\label{fig:NNN}
\end{figure}




\section{Methods}

\subsection {Caging order parameters}
\label{sec:definitions_caging}


In the framework of replica glass theory~\cite{parisi2020theory}, the three transitions considered in this study (the melting transition, the Gardner transition, and the liquid to glass transition) are all characterized by {\it caging order parameters}.
Particles diffuse in the liquid phase, but in the glass phase, they are confined  in  cages formed by nearby  particles.
Thus the average cage size $\Delta$ 
can be used as an order parameter to 
identify glass and melting transitions: 
$\Delta$ is small and finite in glasses (if activations are neglected as in mean-field theories~\cite{parisi2020theory}),  
while in liquids, $\Delta \to \infty$ in the thermodynamic limit.

The order parameter for the Gardner transition is much more complicated. 
In the marginally stable glass phase, particle cages split into multiple hierarchical  sub-cages.
According to the mean-field theory~\cite{charbonneau2017glass, parisi2020theory}, 
the two phases around a Gardner transition are distinguished by, not a single scalar, but a probability distribution function  $P(\Delta_i)$ of single-particle cage size $\Delta_i$.
However, in three-dimensional systems, as shown in this paper and in Ref.~\cite{li2021determining}, it is more practical to look 
at the distribution $P(\chi_i)$ of   single-particle caging susceptibility $\chi_i$,
 which displays single and double peaks in stable and marginal glasses respectively.

\subsection{Preparation of input data using the replica construction}

Practically, the caging order parameters are measured using the replica construction (let us consider $N_{\rm r}$ replicas of the same sample)~\cite{ parisi2010mean, charbonneau2017glass, parisi2020theory, berthier2016growing}. 
The replicated configurations 
are sampled from a given distribution $P(C)$ depending on the state of the system.
In liquids, $P(C)$ is the equilibrium distribution and $C$ can be any microscopic state in the ensemble. In glasses, there are multiple meta-stable states $\mathcal{C}_{\rm MS}$, and different samples may belong to different meta-stable states. Thus the replicas of a given sample should be constructed from the conditional probability distribution $P(C| \mathcal{C}_{\rm MS})$, where $\mathcal{C}_{\rm MS}$ is the meta-stable state that the given sample belongs to. From a dynamical point of view, it means that the same particle from different replicas should be confined in the same cage, but it is allowed to vibrate freely inside the cage. The purpose of replica construction is to translate caging dynamics to a static distribution of replicas. With this setup, the properties of cages
can be obtained by analyzing the ensemble of replicas. 

The design of our input data and network architecture is inspired by the replica construction
 (see Fig.~\ref{fig:NNN}).
While the original system has $N$ particles, the replicated system consists of $N$ ``molecules" at $\{\vec{R}_i\}$, where $i=1 \ldots N$, and each  molecule is formed by  $N_{\rm r}$ replicated ``atoms" at coordinates $\vec{R}_i=\{\vec{r}_i^A \}$, where $A=1 \ldots N_{\rm r}$ is the replica index. Note that the spatial distribution of atoms  inside a molecule represents dynamical rather than structural information: 
in the glass state, this distribution describes how the particle in the original unreplicated system vibrates in its cage.
Although in principle the replica construction is equivalent to brute-force dynamical simulations, the former can facilitate the sampling, which is a significant advantage considering the slow dynamics in the glass state~\cite{charbonneau2015numerical, berthier2016growing}.

In ML, it is conventional to perform additional treatment on the raw coordinate data $\{\vec{R}_i\}$; without imposing inherent physical constraints, a necessary demand for big  training data set would inevitably result in the inefficiency and inaccuracy in statistics~\cite{noe2020machine}.  We thus incorporate the translational and  rotational
symmetries of the physical system, and 
use as our input data  the squared inter-atomic distances,
$\{|\vec{r}_i^{AB}|^2\}$, where $\vec{r}_i^{AB} = \vec{r}_i^A - \vec{r}_i^B$ (see Fig.~\ref{fig:NNN}). For the Gardner transition, we use the normalized quantity, $\{u_{AB}^i\}$, for the input, 
where $u_i^{AB} = \frac{|\vec{r}_i^{AB}|^2} {\langle |\vec{r}_i^{AB}|^2 \rangle_{\rm r}} -1$, with $\langle \cdots \rangle_{\rm r}$ representing the average over $M_{\rm r} = N_{\rm r}(N_{\rm r} -1)/2$ pairs of replicas $A$ and $B$. 
In short, for each sample, at a series of target temperatures $\hat{T}$, we prepare a one-dimensional array $\{ \vec{V}_i \}$ of $M_{\rm r} \times N$ elements as the input data to the neural network, where $\vec{V}_i = \{|\vec{r}_i^{AB}|^2 \}$ for melting and glass transitions, and $\vec{V}_i = \{ u_i^{AB} \}$ for the Gardner transition.

For each transition, 
we perform 
 MD simulations to obtain $N_{\rm s}^{\rm total}$ independent samples, and
divide them into three non-overlapping data sets for the purposes of training ($N_{\rm s}^{\rm train}$ samples), validation ($N_{\rm s}^{\rm valid}$ samples) and prediction ($N_{\rm s}^{\rm pred}$ samples), where $N_{\rm s}^{\rm total}=N_{\rm s}^{\rm train}+N_{\rm s}^{\rm valid}+N_{\rm s}^{\rm pred}$. Below we explain in detail how the input data to NNN are prepared, for the three types of transitions, accordingly.

\subsubsection{Melting transition}

The initial state of each sample is an equilibrium liquid configuration at $\hat{T}_{\rm g} = 0.033$, generated by the swap algorithm. We then make $N_{\rm r}$ replicas of each sample, and decompress them  using the Lubachevsky-Stillinger algorithm~\cite{lubachevsky1990geometric} with a negative compression rate $\Gamma=-10^4$, to  $N_{\hat{T}}$ different target temperatures in a window $\hat{T} \in (0.033, 0.123$). 
The replicas share the same initial configuration at $\hat{T}_{\rm g}$, but are assigned to different initial particle velocities drawn from the Maxwell-Boltzmann distribution.  
After decompression,  the squared inter-atomic distances, $\{ |\vec{r}_i^{AB}|^2\}$,
are computed and  used as the input to NNN. For the $N_{\rm r}$ replicas,  there are in total $M_{\rm r}=N_{\rm r}(N_{\rm r}-1)/2$ pairs can be formed, $(A,B) = (1,1), (1,2), ... (N_{\rm r}-1, N_{\rm  r})$. The vector of  $\{  |\vec{r}_i^{AB}|^2 \} =  \{  |\vec{r}_i^{11}|^2,  |\vec{r}_i^{12}|^2, \ldots,  |\vec{r}_i^{N_{\rm r}-1, N_{\rm  r}}|^2 \}$ 
is fed into the $i^{\rm th}$ small network at the first level (see Fig.~\ref{fig:NNN}). The complete input data is a vector of $N \times M_{\rm r}$ element, 
$\{ \{ |\vec{r}_1^{AB}|^2 \}, \{ |\vec{r}_2^{AB}|^2\}, \ldots, \{ |\vec{r}_N^{AB}|^2\}\}$.
For the melting transition, we use $N_{\rm s}^{\rm total} = 2400 $, $N_{\rm s}^{\rm train} = 1800$, $N_{\rm s}^{\rm valid} = 300$, $N_{\rm s}^{\rm pred} = 300$, and $N_{\hat{T}} = 35 $.

\subsubsection{Gardner transition}

The procedure is similar to the one for the melting transition, except that the compression rate $\Gamma = 10^{-4}$ is positive.
For the input data, instead of 
$\{ |\vec{r}_i^{AB}|^2\}$, we use the normalized quantities
 $\{ u_i^{AB} = \frac{|\vec{r}_i^{AB}|^2} {\langle |\vec{r}_i^{AB}|^2 \rangle_{\rm r}} -1 \}$.
  It turns out that such a simple pre-treatment can efficiently improve the performance of our machine learning model (see \ESI\ Sec.~S7).  The following parameters are used: $N_{\rm s}^{\rm total} = 2400 $, $N_{\rm s}^{\rm train} = 1800$, $N_{\rm s}^{\rm valid} = 300$, $N_{\rm s}^{\rm pred} = 300$, and $N_{\hat{T}} = 23$ (target $\hat{T}$ is chosen in a window $\hat{T} \in (0, 0.033$)).

\subsubsection{Glass transition}

The initial state of each sample is a dilute liquid configuration at  $\varphi = 0.2$.  
Each sample is compressed to a target density $\varphi$ (alternatively one can set a target temperature $\hat{T}$, or pressure $\hat{P}$) with a fixed rate $\Gamma$, using the Lubachevsky-Stillinger algorithm. We choose in total $N_{\hat{T}}$  different  target densities in a window $\hat{T} \in (0.002, 0.268)$.
Once the target $\hat{T}$ is reached, we stop compression and make $N_{\rm r}$ replicas of each sample. These replicas share the same particle positions, but the particle velocities are independently reset according to the Maxwell–Boltzmann distribution. After making $N_{\rm r}$ replicas, we reset simulation time $t$ to zero, and then 
perform constant volume ($\Gamma = 0$) MD simulations. Note that the $N_{\rm r}$ replicas evolve independently because they are assigned to different  velocities at $t=0$. We then collect configurations of these replicas at $t>0$, and compute  $\{ |\vec{r}_i^{AB}|^2\}$.
For the data presented in this study, we have used $\Gamma = 10^{-3}$, $N_{\rm s}^{\rm total} = 1152$, $N_{\rm s}^{\rm train} = 864$, $N_{\rm s}^{\rm valid} = 144$, $N_{\rm s}^{\rm pred} = 144$, and $N_{\hat{T}} = 29 $.
 Note that this procedure is  different from the one explained above for melting  and Gardner transitions: 
 for the glass transition, the replicas are created after compression, while for melting and Gardner transitions, they are created before decompression/compression.
 


\subsection{Machine learning method}

\color{black}


The structure of the input data and that of the two-level NNN are naturally matched. 
Each small network at the first level is responsible for a molecule, whose function is to map the input vector $\vec{V}_i$  to a scalar $q_i$.
\magenta{Physically, the machine learned parameter $q_i$ are intrinsically correlated to characteristic caging
order parameters (see below).}
{ The mapping between the input data and the caging order parameter is independent of the choice of activation functions for both melting and Gardner transitions.}
The  fully connected feed-forward neural network (FNN) at the next level takes 
$\{ q_i \}$ of the whole system as the input and makes a phase classification. 
\magenta{Note that the present  algorithm  can autonomously capture caging features from replicated configurations, and in the meanwhile realize the phase classification. It is thus
prominently distinguished from the previous study on the Gardner transition~\cite{li2021determining}, where the caging features were judiciously  pre-designed as the input to, rather than provided as the output from,  the ML algorithm.}




For supervised phase classification, one needs to label samples  prior to training~\cite{Carrasquilla2017Machine}.
This is realized by designing a {\it blanking window} $[\hat{T}_2, \hat{T}_1]$, out the range of which samples are labeled only; namely, we assign training  samples at $\hat{T}>\hat{T}_1$ to phase I and those at $\hat{T}<\hat{T}_2$ to phase II. 
\magenta{In principle, the results of our algorithm  should not rely on the blanking window, fixed by hyperparameters  $\hat{T}_1$ and $\hat{T}_2$ (ML results should not depend on  hyperparameters); in other words, any predictions depending on such hyperparameters would be ``unphysical" and need to be addressed with care (see below). }


Once well trained, the algorithm makes 
a prediction of the probability $P(\hat{T}, N)$ (or $1-P(\hat{T}, N)$) of a $N$-particle system belonging to phase I (or phase II), at any given temperature $\hat{T}$, calculated from test samples. The data are then fitted to an error function,
\beq
P(\hat{T}, N) = \frac{1}{2} + \frac{1}{2}\erf \left\{\left[\hat{T} - \hat{T}_{\rm c}(N)\right]/w(N) \right\},
\label{eq:ML_fitting}
\eeq
to estimate both the transition temperature $\hat{T}_{\rm c}(N)$ and the width $w(N)$ of transition regime (see Fig.~\ref{fig:NNN}, $\hat{T}_{\rm c} = \hat{T}_{\rm m}$, $\hat{T}_{\rm G}$ and $\hat{T}_{\rm g}^{\rm ML}$ for melting, Gardner and glass transitions respectively). {Additional details on the ML method is provided in \ESI~Sec.~S3}.


\subsection{Finite-size analysis method}

\color{black}
To determine the order of phase transition, we perform a finite-size scaling analysis to the ML output~\cite{Carrasquilla2017Machine,li2021determining}, 
\beq
P(\hat{T}, N) =  \mathcal{P}\left( |\hat{T}-\hat{T}_{\rm c}| N^b \right).
\label{eq:P_finite_size_scaling}
\eeq
Here (i) $b = 1$ for a standard first-order phase transition without disorder, (ii) $b=1/2$ for a first-order phase transition with disorder, and (iii) $b=1/d\nu$ for a second-order phase transition.
{In addition,} $d=3$ is the dimensionality and $\nu$  the critical exponent for the divergence of correlation length.
In general, the value of $b=1/d\nu$ in (iii)
does not equal to $1$ or $1/2$, 
and thus by measuring  $b$, one can distinguish between the above-mentioned three kinds of phase transitions. In practice, $b$ is estimated by fitting the transition width data $w(N)$ to,
\beq
w(N) = w_0 N^{-b},
\label{eq:ML_w_scaling}
\eeq
which can be derived from Eq.~(\ref{eq:P_finite_size_scaling}) ($w_0$ is a constant prefactor, see Fig.~\ref{fig:NNN}).

\begin{figure}[th]
\centerline{\includegraphics[width=0.9\columnwidth]{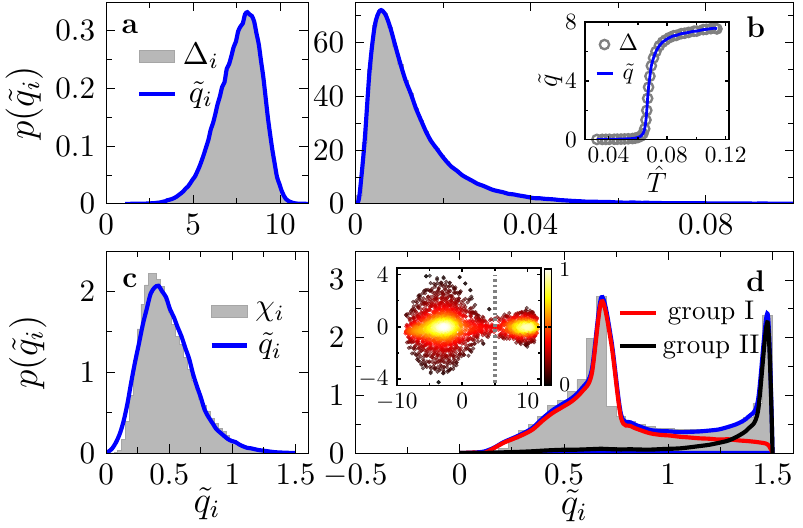}} 
\caption{Learning caging order parameters in (a-b) melting  and (c-d) Gardner  transitions
($N=125$ particles).
For the melting transition, the two distributions $P(\Delta_i)$ and $P(\tilde{q}_i)$ nearly coincide in both (a) liquid ($\hat{T} = 0.13$) and (b) glass ($\hat{T} = 0.03$) phases. The Pearson correlation coefficient between $\Delta_i$ and $q_i$ is  $r = 1.0$ in both cases.  (b-inset) Averaged $\Delta(\hat{T})$ and $\tilde{q}(\hat{T})$ around melting. 
For the Gardner transition, 
the distributions $P(\chi_i)$ and  $P(\tilde{q}_i)$ are plotted in (c) the stable glass phase ($\hat{T} = 0.03$)
and (d) the marginally stable glass phase ($\hat{T} = 10^{-5}$), with $r= 0.96$ 
and $r= 0.99$ respectively. 
(d-inset) Two groups of samples in the marginal phase  visualized by the t-SNE method, where each point represents a sample  and the color bar indicates normalized density of data points. 
The vertical dashed line indicates the estimated boundary between two groups, whose $P(\tilde{q}_i)$ are plotted in the main panel.
}
\label{fig:caging}
\end{figure}

\section{Results}
\subsection{Melting transition}
We first apply our method to the melting of ultra-stable glasses. We find that, once well trained, the machine learned parameter $q_i$ is perfectly correlated to the physically defined caging order parameter $\Delta_i = \langle \left| \vr_i^A - \vr_i^B \right |^2 \rangle_{\rm r}$, which characterizes the cage size of particle $i$~\cite{charbonneau2015numerical, berthier2016growing, parisi2020theory} { (see Fig.~\ref{fig:caging}(a-b) and Fig.~\ref{fig:tilde_q})}.
The calculated  Pearson correlation coefficient,  $r \equiv  \frac{\sum_i (q_i - q) (\Delta_i - \Delta)}{\sum_i (q_i - q)^2 \sum_i (\Delta_i - \Delta)^2} = 1.0$,  independent of $\hat{T}$, where $q = \overline{\langle q_i \rangle_i}$ and $\Delta = \overline{\langle \Delta_i \rangle_i}$, with $\langle \cdots \rangle_{i}$ and $\overline{\cdots}$ being {{ averages} over particles and samples. 
The two parameters can be quantitatively matched by a linear rescaling, $\tilde{q}_i \equiv  c_1 q - c_2$, where $c_1$ and $c_2$ are constants for $i$ and $\hat{T}$ but dependent on the initial parameterization  in each training (see { Fig.~\ref{fig:tilde_q}}).
The agreement is demonstrated in Figs.~\ref{fig:caging}(a-b) by comparing the distributions $p(\tilde{q}_i)$ and $p(\Delta_i)$, both before and after melting, and the average values $\tilde{q}(\hat{T})$ and $\Delta(\hat{T})$ at any $\hat{T}$.

{ To determine the parameters $c_1$ and $c_2$, we collect the pairs of $q_i$ and $\Delta_i$ 
of all $N$ particles in $N_{\rm s}$  samples at $N_{\hat{T}}$ different temperatures (covering both phases), and perform a linear fitting (see Fig.~\ref{fig:tilde_q}(a) for an example). We find that the values of $c_1$ and $c_2$ are non-deterministic, but the agreement between the rescaled predictions and the physically defined caging order parameters is very robust (Fig.~\ref{fig:tilde_q}(b)). It implies that the evolution of caging order parameter is correctly captured by NNN, but it is unnecessarily to fix the rescale parameters $c_1$ and $c_2$ for the purpose of phase classification.}


\begin{figure}[ht]
\centerline{\includegraphics[width=0.8\columnwidth]{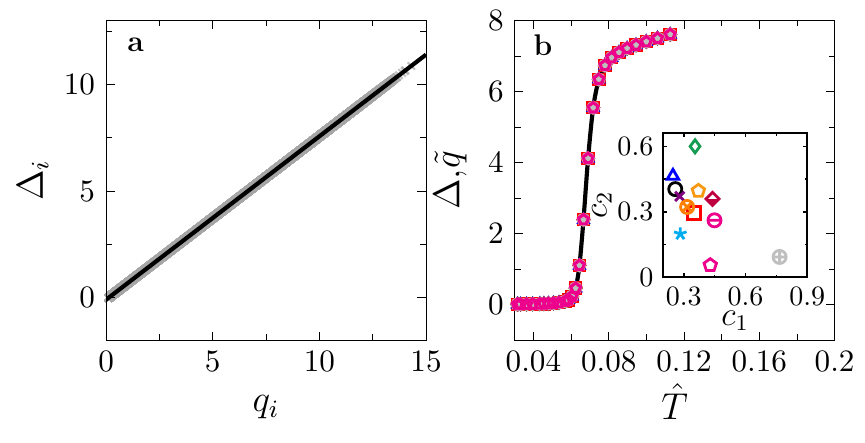}} 
\caption{{ {Rescaling the predicted caging order parameters.}
(a) Determination of $c_1$ and $c_2$ for the melting transition. The data points of  $q_i$ and $\Delta_i$ are collected from $N=125$ particles in $N_{\rm s}=64152$ samples in the temperature window $\hat{T} \in (0.033,0.123)$. The line represents the best fitting, $\Delta_i  = c_1 q_i - c_2$, where $c_1$ and $c_2$ are $\hat{T}$-independent fitting parameters. 
With the fixed $c_1$ and $c_2$, we define $\tilde{q}_i \equiv c_1 q - c_2$.
(b) 
Comparison of the averaged  prediction $\tilde{q} = \frac{1}{N} \overline{\sum_i \tilde{q}_i}$ (points) and the physical order parameter $\Delta = \frac{1}{N} \overline{\sum_i \Delta_i} $ (line),
at different 
$\hat{T}$
around melting. 
Note that $c_1$ and $c_2$ are fixed for each run of training, but vary for different runs (the initial values of network parameters are the same, but the training procedure is stochastic).
We perform 12 independent runs, and plot the obtained $c_1$ and $c_2$ in the inset.}}
\label{fig:tilde_q}
\end{figure}


\begin{figure}[th]
\centerline{\includegraphics[width=0.838\columnwidth]{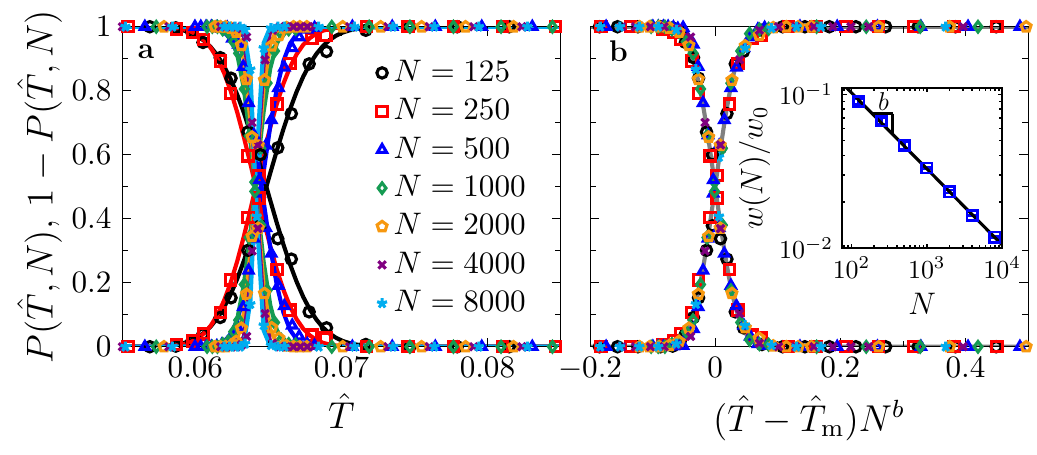}} 
\caption{{ {Finite-size analysis of the machine learning results of the melting transition.} 
(a) Machine learning results $P(\hat{T}, N)$  and $1-P(\hat{T}, N)$
as functions of  $\hat{T}$  and (b) $(\hat{T} -\hat{T}_{\rm m})N^b$, where  $b=0.496(5)$.
The data points are fitted to an error function (lines),
$P(\hat{T}, N) = \frac{1}{2} + \frac{1}{2}\erf \left\{\left[\hat{T} - \hat{T}_{\rm m}(N)\right]/w(N) \right\}$, where the transition temperature $\hat{T}_{\rm m}(N)$ (see 
{Fig.~\ref{fig:ML}(a)}) and the width $w(N)$ (see b-inset) are two fitting parameters. 
The exponent $b=0.496(5)$ is obtained 
by fitting the $w(N)$ data to the scaling $w(N) = w_0 N^{-b}$.
The data for the same $N$ are represented by the same color in both panels. }
}
\label{fig:ML_melting}
\end{figure}

{ The machine learning results $P(\hat{T}, N)$ of the melting transition are fitted to Eq.~(\ref{eq:ML_fitting}), which gives the melting temperature $\hat{T}_{\rm m}(N)$ and the width of transition regime $w(N)$ (see Fig.~\ref{fig:ML_melting}). 
The determined melting transition temperature $\hat{T}_{\rm m} \approx 0.064$ 
 is} consistent with the estimation  from the peak of density fluctuation $\chi_{\varphi}(\hat{T})$ (see \ESI~Fig.~S2(a)
and Fig.~\ref{fig:ML}(a)). The transition width data $w(N)$ are fitted to Eq.~(\ref{eq:ML_w_scaling}), giving $b = 0.496(5)$.
The result $b \simeq 1/2$ suggests that the melting of ultra-stable glasses is a first-order transition with disorder, consistent with the discontinuity on the  equation of state (EOS) $\hat{T}(\varphi)$ around $\hat{T}$(see Fig.~S1 of \ESI).
Note that $b \simeq 1/2$ for the melting transition can be confirmed by the finite-size analysis of the physically-defined density susceptibility 
$\chi_{\varphi} /N^a = \mathcal{X}\left( |\hat{T}-\hat{T}_{\rm c}| N^b \right)$,
with an additional parameter $a\approx 1$ (see \ESI~Fig.~S2(b)).
This scaling function suggests the existence of 
two susceptibilities, a {\it disconnected} one, $\chi_{\rm dis} = \chi_{\varphi} \sim N^a \sim N$, and a {\it connected} one $\chi_{\rm con} \sim d\mathcal{X}/d\hat{T} \sim  N^b \sim N^{1/2}$, related via, $\chi_{\rm dis} \sim \chi^2_{\rm con}$. 
Such a relationship seems to be ubiquitous in many transitions in disordered systems:  
it presents also around the critical point in the
random field Ising model~\cite{gofman1993evidence, nattermann1998theory}, in the  yielding of amorphous solids~\cite{ozawa2018random}, and near the critical point in a replicated glass-forming model with coupling~\cite{berthier2015evidence}.

{The output $P(\hat{T}, N)$ represents the fraction of liquid samples identified by ML among the entire set of samples provided for prediction, for the given $N$ and $\hat{T}$. 
Assuming that the system is self-averaging, i.e., a single large system is sufficient to represent the whole ensemble, one may also interpret $P(\hat{T}, N)$ as the fraction of liquid-like particles in the sample. However, the inverse problem is not straight-forward (if not impossible), i.e., using a physical approach to find these  liquid-like particles such that their probability is identical to $P(\hat{T}, N)$. 
Around the melting transition, it is not easy to define a sharp boundary between liquid-like and solid-like particles based on physical quantities such as the cage size. Furthermore, the property of self-averaging may break down in small systems with quenched disorder.}


\begin{figure}[th]
\centerline{\includegraphics[width=1.\columnwidth]{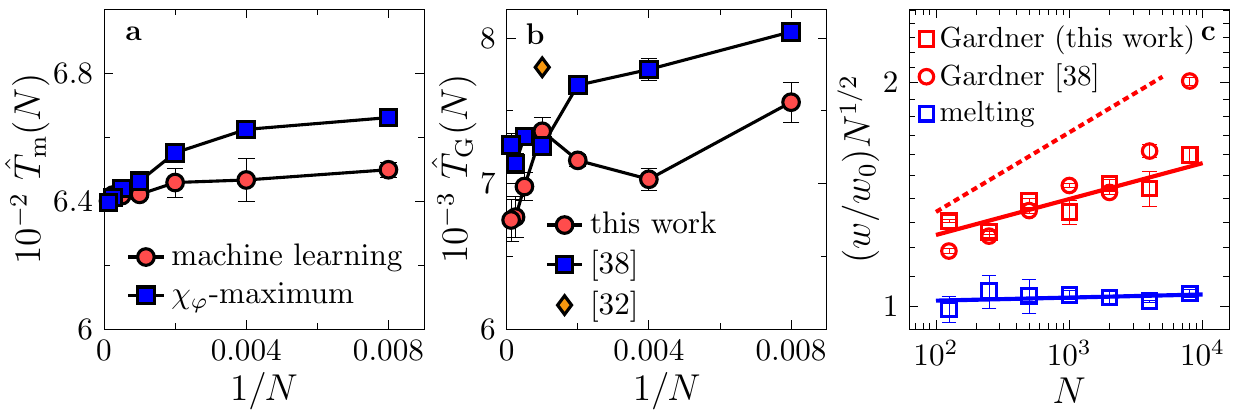}} 
\caption{Machine learned results of melting and Gardner transitions. 
(a) Melting temperature $\hat{T}_{\rm m}(N)$ of systems with different sizes $N$, obtained from ML and the maximum of  density susceptibility $\chi_\varphi$. 
(b) Machine learned Gardner transition temperature $\hat{T}_{\rm G}(N)$ obtained in  this work and in Ref.~\cite{li2021determining}, as well as $\hat{T}_{\rm G}(N=1000) \approx 0.0078$ by a physical method in Ref.~\cite{berthier2016growing}.
(c) Finite-size scaling  of machine learned transition width $w(N)$ of both melting and  Gardner transitions.
The solid lines represent Eq.~(\ref{eq:ML_w_scaling}), where $b=1/2$ for the blue, and $b=1/d\nu = 0.45(1)$ ($\nu = 0.74(2)$) for the red. The red dashed line indicates the theoretical result $\nu = 0.85$~\cite{charbonneau2017nontrivial}.
Error bars represent the standard error of the mean in all figures.
}
\label{fig:ML}
\end{figure}


\subsection{Gardner transition}
Compared to the melting transition, the Gardner transition has much more subtle signatures -- e.g.,  
around the transition, the  $\hat{T}(\varphi)$ EOS does not display any jump or kink~\cite{berthier2016growing}.
Figures~\ref{fig:caging}(c-d) show that the machine learned $q_i$ is highly correlated to the single particle caging susceptibility $\chi_i \equiv \langle (u^{AB}_i)^2 \rangle_{\rm r} - \langle u^{AB}_i \rangle_{\rm r} ^2$   characterizing the fluctuation of cage size, i.e., $\tilde{q}_i \simeq \chi_i$.  The distribution $p(\tilde{q}_i)$ shows a characteristic two-peak feature in the marginally stable phase ($\hat{T}< \hat{T}_{\rm G}$), similar to the behavior of  $p(\chi_i)$ observed previously~\cite{li2021determining}. 
The two peaks in $p(\tilde{q}_i) \simeq p(\chi_i)$ demonstrate the existence of 
two dynamical caging modes
in the marginally stable phase: 
particles with a larger/smaller $\chi_i$ should have a more heterogeneous/homogeneous cage shape. This difference on single particle caging behavior is also reflected at the system level. 
Applying the unsupervised t-distributed stochastic neighbor embedding (t-SNE)  method to $\{\tilde{q}_i\}$, we identify two groups of samples in the marginal phase, containing mostly heterogeneous/homogeneous caging particles respectively (see  Fig.~\ref{fig:caging}(d)). 
ML fails to identify  the distribution of cage sizes $p(\Delta_i)$,   
suggested by the mean-field replica theory~\cite{charbonneau2014fractal, parisi2020theory, berthier2016growing}, as features for phase classifications
(see {Fig.~S10} of \ESI)
, which shows the complexity of learning caging order parameters in finite dimensions.

The finite-size scaling of $P(\hat{T}, N)$ and $w(N)$ for the Gardner transition satisfy  Eqs.~(\ref{eq:P_finite_size_scaling}) and~(\ref{eq:ML_w_scaling}) as well { (see Fig.~\ref{fig:Gardner})}. The estimated transition temperature $\hat{T}_{\rm G}(N)$  is close to previously reported values by a physical approach~\cite{berthier2016growing} and ML using handcrafted descriptors~\cite{li2021determining}  (see Fig.~\ref{fig:ML}(b)).


\begin{figure}[th]
\centerline{\includegraphics[width=0.838\columnwidth]{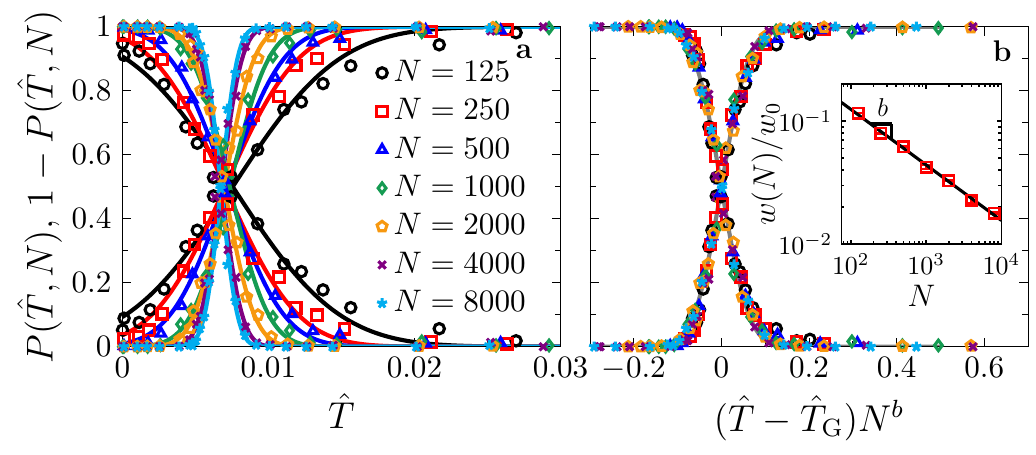}} 
\caption{{ {Finite-size analysis of the Gardner transition.}
(a) Output from machine leaning, $P(\hat{T}, N)$  and $1-P(\hat{T}, N)$, as functions of  $\hat{T}$ and (b) $(\hat{T}-\hat{T}_{\rm G})N^b$, for a few different $N$. The data points are fitted to an error function (lines),
$P(\hat{T}, N) = \frac{1}{2} + \frac{1}{2}\erf \left\{\left[\hat{T} - \hat{T}_{\rm G}(N)\right]/w(N) \right\}$, where the transition temperature $\hat{T}_{\rm G}(N)$ (
{Fig.~\ref{fig:ML}(b)}) and the width $w(N)$ (b-inset) are two fitting parameters. The exponent $b = 1/d\nu = 0.45(1)$ is obtained by fitting the $w(N)$ data to the scaling $w(N) = w_0 N^{-b}$.}
}
\label{fig:Gardner}
\end{figure}

Figure~\ref{fig:ML}(c) emphasizes the difference on the finite-size scaling of $w(N)$ in both melting and Gardner transitions. Such a plot can be used to identify the order of phase transitions: a first-order transition (melting) with disorder corresponds to $b=1/2$ in Eq.~(\ref{eq:ML_w_scaling}) and thus to a horizontal line in Fig.~\ref{fig:ML}(c), while a second-order transition (Gardner) should correspond to a line with a non-zero, $\nu$-dependent slope.  The second-order nature of the Gardner transition is supported by the two-peak behavior of $P(\chi_i)$  in Fig.~\ref{fig:caging}(d), which is a universal feature of the probability distribution of order parameters
below the critical temperature according to the Landau theory. 
 Based on the relation $b = 1/d\nu$, we obtain  the critical exponent 
$\nu = 0.74(2) $,  close to the previous data, $\nu = 0.78(2)$ from  a ML numerical  analysis~\cite{li2021determining} and $\nu = 0.85$ from a  field-theory calculation~\cite{charbonneau2017nontrivial}.  
 In addition, the method can be applied to standard first-order ($b=1$) and second-order phase transitions without disorder, such as those in the Ising model (see {\color{blue}Sec.~S5} of \ESI).

\subsection{Glass transition}
 Naturally, one could ask whether the above method can be applied to the liquid to ordinary glass transition. 
 { The difference between the two kinds of glasses, ordinary ($\hat{T}_{\rm g} \approx 0.047$) and ultra-stable ($\hat{T}_{\rm g} \approx 0.033$), is revealed by the hysteresis in their EOSs (see Fig~\ref{fig:glass_transition_caging}(a)).}
 Around $\hat{T}_{\rm g}$, the average cage size $\Delta$ of ordinary glasses 
depends strongly
on the observation time $t$ collapsed after compression, because particle hopping between cages is non-negligible {(see \ESI~Fig.~S12)}.
Nevertheless, for a fixed $t$, the evolution $\Delta(\hat{T})$ can still be accurately captured by the machine learned $\tilde{q}(\hat{T})$  { (see Fig~\ref{fig:glass_transition_caging}(b))}.

\begin{figure}[th]
\centerline{\includegraphics[width=0.891\columnwidth]{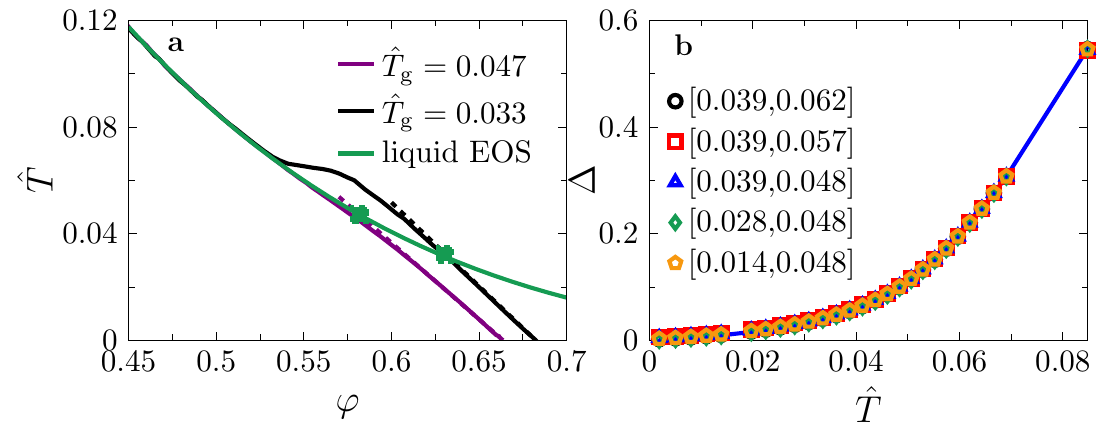}} 
\caption{{ {Learning the caging order parameter of the glass transition.} (a) EOSs for $\hat{T}_{\rm g} = 0.033$  and  $\hat{T}_{\rm g} = 0.047$ ($N=500$).
The glass transition temperature $\hat{T}_{\rm g}$ is practically 
defined as
the intersection of the liquid EOS (green line) and the linearly fitted glass EOS (dashed line).  
(b) The average cage size $\Delta$ (measured with $t=1$) as a function of $\hat{T}$ (line),  for ordinary glasses ($\hat{T}_{\rm g} = 0.047$), together with the machine learned $\tilde{q}(\hat{T})$ using different blanking windows $[\hat{T}_2, \hat{T}_1]$ (points).}
}
\label{fig:glass_transition_caging}
\end{figure}


\begin{figure}[th]
\centerline{\includegraphics[width=0.457\columnwidth]{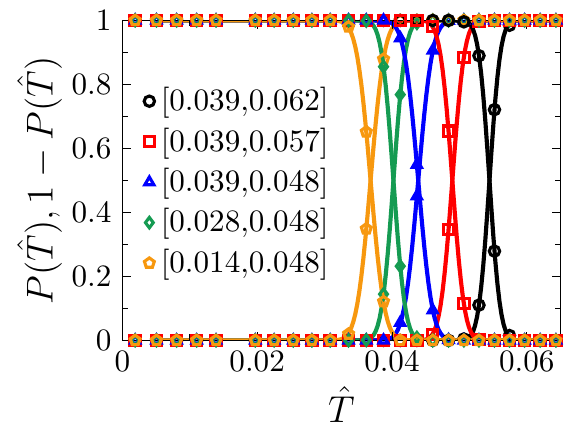}} 
\caption{{ {Non-uniqueness of the glass transition point determined by machine learning.}
Output from machine leaning, $P(\hat{T})$  and $1-P(\hat{T})$, as functions of $\hat{T}$, using a few different blanking windows (for $N=500$ systems,  $t=1$). The data points are fitted to an error function (lines),
$P(\hat{T}) = \frac{1}{2} + \frac{1}{2}\erf \left\{\left[\hat{T} - \hat{T}^{\rm ML}_{\rm g}\right]/w \right\}$.
The estimated $\hat{T}^{\rm ML}_{\rm g}$ is plotted in {Fig.~\ref{fig:glass}(c)}.}
}
\label{fig:glass_transition_learning}
\end{figure}

Interestingly, the ML algorithm fails to identify a unique ``transition point" { (see Fig.~\ref{fig:glass_transition_learning}).}
The estimated crossover temperature $\hat{T}^{\rm ML}_{\rm g}$ 
relies on the labeling of two phases during training, i.e., the blanking window (see Fig.~\ref{fig:glass}(c)). { {Figure~\ref{fig:glass}(c)} also shows that $\hat{T}^{\rm ML}_{\rm g}$ is nearly identical to the center of blanking window, $\hat{T}_{\rm center}$, suggesting that $\hat{T}^{\rm ML}_{\rm g}$ simply separates low-density and high-density states, whose definitions are preset artificially by the blanking window.}
In contrast, the estimated transition temperatures of melting and Gardner transitions are independent of such algorithm hyperparameters (see Figs.~\ref{fig:glass}(a-b) and {\ESI~Figs.~S9 and S11}).
Although the predicted $P(\hat{T}, N)$ would look very similar for both dynamical crossovers and phase transitions (as the one plotted in Fig.~\ref{fig:NNN}), one should perform additional checks on the independence of hyperparameters (such as the blanking window discussed here) in order to confirm a true phase transition.

\begin{figure}[th]
\centerline{\includegraphics[width=0.8\columnwidth]{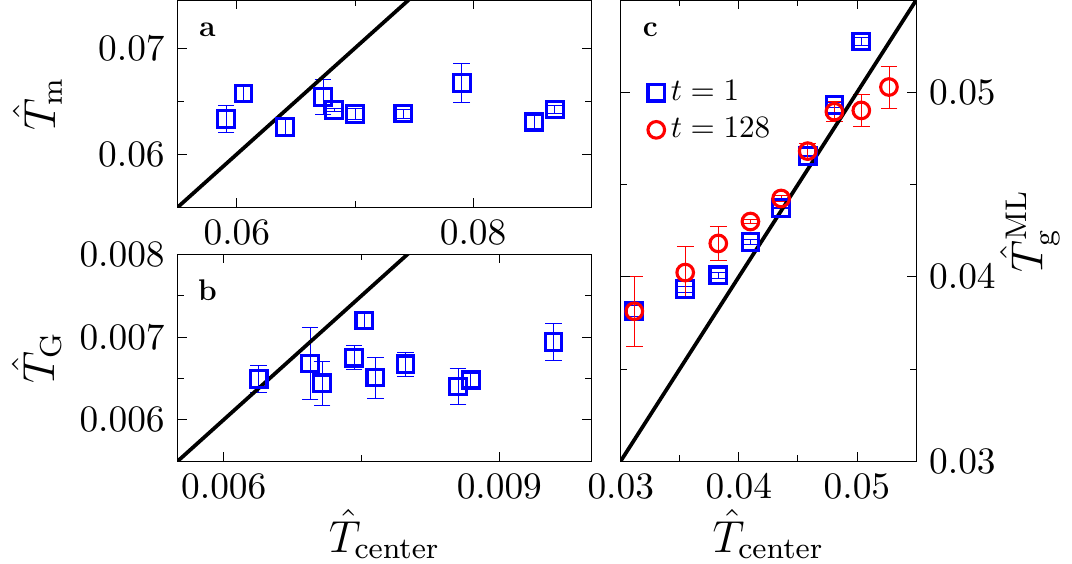}} 
\caption{Dependence of machine learned transition temperatures on the center of blanking window $\hat{T}_{\rm center}$.
(a) The melting temperature $\hat{T}_{\rm m}$ (data obtained for $N=2000$ systems) and (b) the Gardner transition temperature $\hat{T}_{\rm G}$ ($N=8000$) are independent of $\hat{T}_{\rm center}$.
(c) The crossover $\hat{T}^{\rm ML}_{\rm g}$ from liquids to ordinary glasses is non-unique ($N=500$, two different observation  times $t$). 
The { solid} lines represent $y=x$.
}
\label{fig:glass}
\end{figure}



\section{Conclusion}
\magenta{We demonstrate that ML algorithm can identify caging features of glasses, based on which, phase classifications can be accomplished.}  {In general, the input vector to the NNN, $\vec{V}_i = \{v_i^1, v_i^2, \cdots, v_i^{M_{\rm r}}\}$, represents a probability distribution function $p(v_i)$ describing certain features of  particle $i$. The function of the front-connected small networks is to map the distribution  $p(v_i)$ to a scalar order parameter $q_i$, which can be either the mean (as in the case of melting and glass transitions) or the variance of $p(v_i)$ (as in the case of Gardner transition). In \ESI~Sec.~S9, via a toy model, we further demonstrate that, the NNN can also identify $q_i$ as the skewness (the third moment) of the input distribution, when the first two moments (mean and variance) do not carry useful information. Without loss of generality, we conclude that, in principle, the presented NNN method is  capable of learning order parameters  encoded in arbitrary-order statistical moments of the input data.}

It would be challenging, if not impossible, to achieve similar classifications using only static information.
The caging dynamics are directly related to the structure of phase space. 
In the language of replica theory, the breakdown of ergodicity in glasses is a result of replica symmetry breaking (RSB)~\cite{parisi2020theory}. 
The two-level structure of NNN can naturally capture  caging order parameters of both the melting 
transition (1-step RSB) and the Gardner transition (full-step RSB). 
Increasing the number of nested levels would be useful for other types of RSB (e.g., 2-step RSB in glasses of bidisperse particles~\cite{ikeda2021multiple}). Finally,  we expect  generalizations of our method to other glassy systems, such as polymers and spin glasses. 

\section*{Author Contributions}
All authors contributed equally to the paper.

\section*{Conflicts of interest} 
The authors declare no competing interests.

\section*{Acknowledgements}

Y.~Jin acknowledges funding from  National Natural Science Foundation of China (Project 11974361, Project 11935002, Project 12161141007, and Project 12047503),
and Chinese Academy of Sciences (the Key Research Program of Frontier  Sciences Grant NO. ZDBS-LY-7017, the Key Research Program Grant NO. XDPB15, and Grant NO. KGFZD-145-22-13).
}
Y.~Jiang acknowledges funding from Project 22073004 supported by National Natural Science Foundation of China.
This work was granted access to the High-Performance Computing Cluster of Institute of Theoretical Physics - Chinese Academy of Sciences and Beihang University.

\bibliography{MLM}

\clearpage

\centerline{\bf \Large Supplementary Information}

\setcounter{figure}{0}  
\setcounter{equation}{0}  
\setcounter{table}{0} 
\setcounter{section}{0} 

\renewcommand\thefigure{S\arabic{figure}}
\renewcommand\theequation{S\arabic{equation}}
\renewcommand\thesection{S\arabic{section}}
\renewcommand\thetable{S\arabic{table}}


\section{Glass model}
\label{sec:model}
The model~\cite{berthier2016equilibrium, berthier2016growing, jin2018stability, jin2017exploring, seoane2018spin} consists of $N=125-8000$ polydisperse hard spheres (HSs), whose diameters are distributed according to a continuous function $P_D (D_{\rm min} \leq D \leq D_{\rm min}/0.45) \sim D^{-3}$. The volume of simulation box is $V$, and periodic boundary conditions are used. 
The system state is characterized by volume fraction $\varphi$ and reduced temperature $\hat{T} =  1/\hat{P} = Nk_{\rm B}T/PV $, where $P$ is the pressure, $\hat{P}$ the reduced pressure, $k_{\rm B}=1$  the Boltzmann constant, and $T=1$  the temperature.
We set the mean diameter  $D_{\rm mean}$ as  unit length, and the particle  mass $m$ as  unit mass. 
Crystallization is suppressed by polydispersity, and will not be discussed in this study.

\begin{figure}[th]
\centerline{\includegraphics[width=0.5\columnwidth]{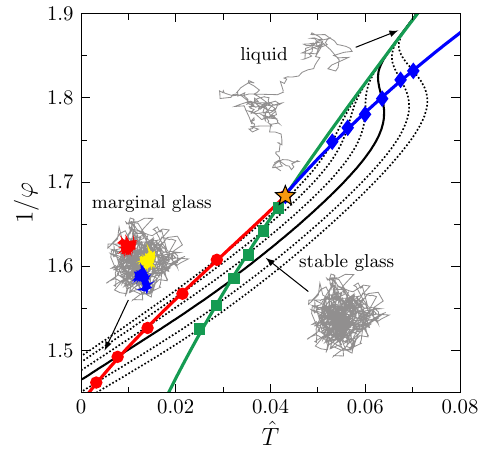}} 
\caption{{Phase diagram of a hard sphere glass model~\cite{berthier2016growing}.} The green, blue and red lines represent the  CS liquid EOS, the melting line that separates liquid and glass phases, and the Gardner line that separates stable and marginally stable glass phases.   The orange star represents the MCT transition point. The system is initially equilibrated at $\hat{T}_{\rm g}$ (green squares), and then  
evolves  following the glass EOSs (dotted black lines) under compression or decompression.  To study melting and Gardner transitions, we focus on the ultra-stable case of $\hat{T}_{\rm g} = 0.033$ (solid black line). 
Typical particle trajectories are plotted to show the diffusive motion in liquids, a confined cage in stable glasses, and the  split of the cage into sub-cages in marginal glasses (the three sub-cages are visualized by trajectories of the same particle in three replicas, which are compressed from the same initial configuration at $\hat{T}_{\rm g}$).}
\label{fig:PD}
\end{figure}

The phase diagram of the model is presented in Fig.~\ref{fig:PD}~\cite{berthier2016growing}.
 Any state of the system is described by two thermodynamic parameters, the volume fraction $\varphi$ and the reduced temperature $\hat{T}$. 
The  Carnahan-{Starling} (CS) equation of state (EOS)~\cite{boublik1970hard} well captures the relationship between $\varphi$  and $\hat{T}$ of liquid states~\cite{berthier2016growing}. The mode-coupling theory (MCT) transition point, $\hat{T}_{\rm MCT} \approx 0.044$ (or $\varphi_{\rm MCT} \approx 0.594$), was estimated in~\cite{berthier2016growing},  below which equilibrium  becomes difficult in ordinary molecular dynamics (MD) simulations. Following convention, the intersection between liquid and glass EOSs is defined as the glass transition point $\{\varphi_{\rm g}, \hat{T}_{\rm g}\}$.
The location of glass transition point 
depends on the compression rate $\Gamma$,
and therefore, not unique. As can be seen in Fig.~\ref{fig:PD}, the EOS of an ultra-stable glass ($\hat{T}_{\rm g} \ll \hat{T}_{\rm MCT}$) displays  enormous overshooting over the liquid EOS and the two EOSs are connected by an abrupt jump.
In contrast, the EOS of a poorly annealed ordinary glass ($\hat{T}_{\rm g} \approx \hat{T}_{\rm MCT}$) merges smoothly to the liquid one.

To study melting and Gardner transitions, we consider ultra-stable glasses. Deeply supercooled liquid states are prepared by using an efficient swap Monte Carlo (MC) algorithm~\cite{berthier2016equilibrium}.
Once the initial states are obtained, we switch to regular MD (without swap) to simulate follow-up dynamics.
These deeply supercooled liquid states have extraordinarily large structural relaxation ($\alpha$-relaxation) time $\tau_\alpha$ in the MD time unit,  much larger than our MD simulation time window.

Two instabilities -- the melting and the Gardner transitions -- occur if one decompresses or compresses an ultra-stable HS glass.
Under decompression, the glass is effectively ``heated" up and eventually melts into a liquid at a melting temperature $\hat{T}_{\rm m} (\hat{T}_{\rm g})>\hat{T}_{\rm g}$. Although glass melting is a non-equilibrium procedure by definition, previous experiments~\cite{swallen2009stable} and simulations~\cite{jack2016melting} showed that, 
this procedure in ultra-stable glasses is very similar to the melting of crystals, which is a first-order phase transition. 
In contrast, the melting of (poorly annealed) ordinary  glasses is a smooth crossover without any discontinuous behavior. 

On the other hand, a {\it Gardner transition}~\cite{gardner1985spin, charbonneau2014fractal} is expected to occur at $\hat{T}_{\rm G} (\hat{T}_{\rm g})<\hat{T}_{\rm g}$ if an ultra-stable HS glass is compressed.
The Gardner transition separates the {\it stable glass} (at $\hat{T}> \hat{T}_{\rm G}$)  and the {\it marginally stable  glass}  (at $\hat{T}< \hat{T}_{\rm G}$)  phases.  
It is predicted to be a second-order phase transition in large dimensions by the mean-field glass theory~\cite{charbonneau2014fractal, parisi2020theory}. Evidence of the Gardner transition in physical dimensions (2D and 3D) has been reported  in a number of simulations~\cite{berthier2016growing,jin2017exploring, liao2019hierarchical} and experiments~\cite{seguin2016experimental,geirhos2018johari,Hammond5714}. A fixed point is found by field-theory calculations, suggesting that the transition survives in low dimensions~\cite{charbonneau2017nontrivial}. 
The existence of a Gardner transition in 3D ultra-stable HS glasses is supported by a recent numerical study, which combines  finite-time-finite-size analyses with machine learning~\cite{li2021determining}. 

We do not study the melting and Gardner transitions in ordinary glasses: The melting of an ordinary glass is nearly reversible to the glass transition, and thus for our purpose, it is sufficient only to consider the latter. The Gardner transition in ordinary glasses is blurred by activated dynamics~\cite{berthier2016growing}; as a result,  we do not expect to observe  critical scalings.

\section{Molecular simulation methods}

\subsection {Molecular dynamics simulations}
\label{sec:MD}
We { use} the Lubachevsky-Stillinger algorithm (event-driven MD) to  simulate compression quench~\cite{lubachevsky1990geometric, skoge2006packing}.
During compression/decompression,
the sizes of all particles are increased/decreased proportionally with a fixed rate  $\Gamma = \frac{1}{2D} \frac{dD}{dt}$. The simulation time is expressed in units of $\sqrt{m  D_{\rm mean}^2/(k_{\rm B}T)}$.
{ 
We simulate each configuration and measure its $\hat{T}$ at a fixed packing fraction $\varphi$.  Then we collect configurations at the desired $\hat{T}$, based on which 
physical quantities are computed. Additional simulation details can be found in Refs.~\cite{berthier2016growing, li2021determining}.}\\


\subsection{Swap algorithm}
\label{sec:swap}
The swap algorithm simulates artificial dynamics that can efficiently accelerate reaching equilibrium ~\cite{berthier2016equilibrium}.
At each swap MC step, two  particles are randomly picked, and swapped if they do not overlap with neighbor particles at the new positions. While the dynamics are unrealistic, the final configurations in equilibrium are equivalent to those generated by standard MD and MC simulations.

\section{Machine learning methods}
\label{sec:ML}
\subsection{Architecture of the artificial neural network}

The  nested neural network (NNN)  comprises two levels of networks, which  in general can have  different structures (see Fig.~1). There are $N$ duplicated small networks at the first level,  each of which extracts the latent caging features of one single particle. 
The small network has only one hidden layer, besides the input and output layers. Both input and hidden layers have $M_{\rm r}$  neuron nodes, 
and the output layer has a single node. 
The $i^{\rm th}$ hidden node is connected by a single link to the $i^{\rm th}$ input node, and  
 is  activated by the exponential linear unit (ELU) function. 
 { We have checked that the machine learning results do not change with a different type of activation function, such as a tanh function,  for both melting and Gardner transitions.}
 The output node simply takes an average of $M_{\rm r}$ hidden nodes. The $N$ small networks 
 share  the same parameters (weights and bias), and thus there are only $ 2 M_{\rm r}$ free parameters at the first level.  
 {The number of parameters can be further reduced from $ 2 M_{\rm r}$ to 2, considering the permutation symmetry of replicas.}
 In principle, one could choose other architectures (e.g., fully connected feed-forward neural network (FNN)) for  small networks, and set their parameters to be independent. In practice, however, we find that using a small number of  free parameters at the first level can significantly increase the efficiency of the NNN model during training, without losing its compatibility and predictive power. 
 {Generally, the number of free parameters can be minimized by considering the symmetries of the physical system under consideration, while networks with redundant parameters can work equally well if proper regularization  is imposed~\cite{kim2018smallest}.}

The $N$ output nodes of the first-level small networks  are considered  as input nodes of the followed big FNN at the second-level. The FNN has one hidden layer of 128 nodes  activated by ELU functions, and one  output layer of two nodes that provide binary classifications through softmax activation functions.



\subsection{Blanking window}
\label{sec:blanking}

For the  supervised learning of phases, we need to label in advance to which phase a given configuration belongs, during training and validation. A blanking window $[\hat T_2, \hat T_1]$ is introduced to skip the vicinity of a (presumed) transition. 
 Specifically, the following setup is used for the data presented in the main figures: 
 for the melting transition, configurations at $\hat T > \hat T_1=0.083$ and $\hat T < \hat T_2=0.053$ are labeled as liquids and glasses respectively; 
 for the Gardner transition, configurations at 
 $\hat{T} > \hat T_1 = 0.011$ and $\hat{T} < \hat T_2=0.0045$ are labeled as stable and marginal glasses respectively. 
 Note that, for both transitions, the machine learning results do not sensitively depend on the choice of  blanking window (see {Fig.~9}, Secs.~\ref{sec:blanking_melting} and~\ref{sec:blanking_Gardner}). In contrast, the learning results of the glass transition correlates strongly to the blacking window.

\subsection{Training the nested neural network and validating predictions}

A cross-entropy cost function is minimized during training.  The Adam optimizer~\cite{kingma2015adam} is used to implement the stochastic gradient descent method for updating the network parameters. To avoid overfitting, a dropout strategy~\cite{srivastava2014dropout} is used, which randomly skips 20\% hidden nodes at each step.
To augment the training data set, we perform $N_{\rm shuffle}=20-200$ random shuffles of the elements in the input vector,  which is equivalent to randomly ordering particle indexes.  In this way, we expand the training data set to  $N_{\rm s}^{\rm train}\times N_{\hat T}\times N_{\rm shuffle} \sim 10^5$ samples. 
The random shuffling apparently destroys spatial correlations (if there is any) between particles.  However, we find that it does not modify the final predictions noticeably. Validation is performed after each training step, by calculating the cost function for the validation data set. The entire training procedure is terminated when the validation cost function reaches a minimum. Such an early stopping strategy can efficiently avoid overfitting.

\subsection{Making predictions using the nested neural network} 

Once well trained, the NNN can make phase predictions for 
the samples 
in the testing data set. 
For each test sample at a temperature $\hat T$, the NNN
provides an  output value of $0$ or $1$. The arithmetic mean of the output over all test samples gives an estimation of the probability $P$ of the system belonging to a specific phase, and $1-P$ to the other.

To achieve a reliable prediction, 
we independently train our NNN for 10 times (runs) and 
calculate  the mean and the statistical error of the predicted $P$.
For each run, $N_{\rm s}^{\rm train}$ training samples and $N_{\rm s}^{\rm pred}$ prediction samples are randomly selected from $N_{\rm s}^{\rm total}$ samples,
and the remaining $N_{\rm s}^{\rm valid}$ samples are used for validation.

\subsection{Unsupervised classification using t-distributed stochastic neighbor embedding}
\label{sec:tSNE}

We utilize the unsupervised t-distributed stochastic neighbour embedding (t-SNE) method  \cite{tSNE} to group samples in the Gardner phase. 
The input data are the machine detected caging parameter $\{\tilde{q}_i\}$
 of each sample, where $i=1,2,\ldots, N$. 
The algorithm 
conducts a nonlinear dimensionality reduction, which maps each vector $\{\tilde{q}_i\}$ to a point in two dimensions.
In the  two-dimensional space, data points are rearranged  according to their similarities quantified by a t-distribution kernel function (see the inset of Fig.~2(d)).

\section{Finite-size scaling function of the susceptibility}

\begin{figure}[th]
\centerline{\includegraphics[width=0.7\columnwidth]{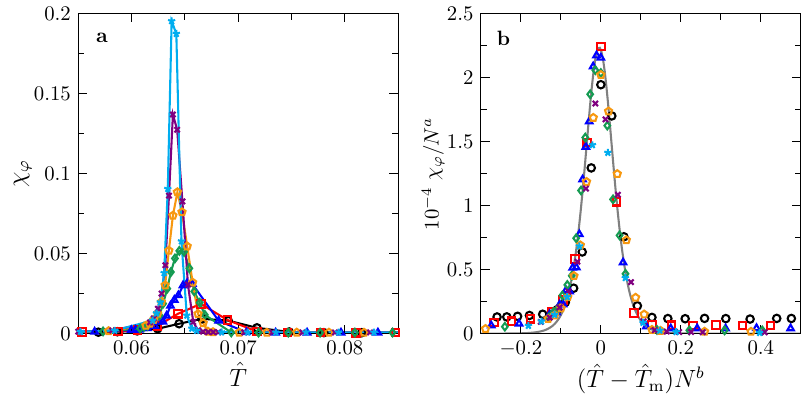}} 
\caption{{Finite-size analysis of the density susceptibility around the melting transition.} 
(a) Density susceptibility $\chi_{\varphi}$ as a function of $\hat{T}$, for  a few different system sizes $N$. (b) Collapse of the $\chi_{\varphi}$ data  according to Eq.~(\ref{eq:chi_finite_size_scaling}). Best collapsing is obtained by setting $a= 0.8$,
$b= 0.5$ 
and $\hat{T}_{\rm c} = \hat{T}_{\rm m} = 0.064$. 
}
\label{fig:melting}
\end{figure}

To determine the order of phase transition, a standard way is to 
 apply a finite-size analysis of data obtained from experiments or simulations.   For example, the fluctuation of order parameter, or the susceptibility, $\chi$, follows a finite-size scaling function around a phase transition, 
\beq
\chi/N^a = \mathcal{X}\left( |\hat{T}-\hat{T}_{\rm c}| N^b \right),
\label{eq:chi_finite_size_scaling}
\eeq 
where $a, b$ are two exponents, $\mathcal{X}(x)$ a scaling function whose concrete form is not important in our discussion, and $\hat{T}_{\rm c}$ the transition temperature. The values of $a$ and $b$ depend on the nature of transition:
(i) For a standard first-order phase transition without disorder,  $a=1$ and $b=1$~\cite{binder1987theory}. An example is 
 the first-order phase transition between positive and negative ferromagnetic phases in the Ising model under an external field.
(ii)  For a first-order phase transition in the presence of disorder, such as the yielding transition~\cite{ozawa2018random} and the melting transition (see Fig.~\ref{fig:melting}) in glasses,
$a=1$ and $b=1/2$. Equation~(\ref{eq:chi_finite_size_scaling}) then results in two susceptibilities, a {\it disconnected} one, $\chi_{\rm dis} = \chi \sim N^a \sim N$, and a {\it connected} one $\chi_{\rm con} \sim d\mathcal{X}/d\hat{T} \sim  N^b \sim N^{1/2}$. 
The two susceptibilities are related via, $\chi_{\rm dis} \sim \chi^2_{\rm con}$, a relation found in the   random field Ising model~\cite{gofman1993evidence, nattermann1998theory}.
(iii) For a second-order phase transition, $a  = b \gamma$ and $ b  = 1/d\nu$, where $d$ is the dimensionality, and $\gamma$ and $\nu$ are the critical exponents for the divergences of susceptibility and correlation length. A standard example is  the second-order phase transition between paramagnetic and ferromagnetic phases in the Ising model without a field. 

Let us examine the finite-size scaling of the
density susceptibility at fixed $\hat{T}$, $\chi_{\varphi} = N \left[ \overline{ \varphi^2 } - \left(\overline{ \varphi } \right)^2 \right] / \left(\overline{ \varphi } \right)^2 $,
around the melting of ultra-stable glasses, where $\overline{\cdots}$ represents the average over samples. 
The density susceptibility
displays a clear dependence on system size $N$, around the melting temperature $\hat{T}_{\rm m} \approx 0.064$ (see Fig.~\ref{fig:melting}(a)). Its finite-size scaling satisfies Eq.~(\ref{eq:chi_finite_size_scaling}), with $a \approx 1$ and $b \approx 0.5$ (see Fig.~\ref{fig:melting}(b)). 
The melting temperature $\hat{T}_{\rm m}(N)$ can be obtained from 
the peak position of  $\chi_{\varphi}(\hat{T})$ (data plotted in  {Fig.~5(a)}).
Note that the caging susceptibility around the Gardner transition suffers from strong finite-size and finite-time effects simultaneously, making a direct analysis difficult (see Ref.~\cite{li2021determining}).

\section{Machine learning phase transitions in the Ising model}
\label{sec:Ising}

\subsection{Machine learning algorithm}


We study the  Ising model in both two and three dimensions to validate the finite-size scaling function Eq.~(2) in a standard equilibrium system. Machine learning is carried out using TensorFlow~\cite{abadi2016tensorflow}. 
Following~ \cite{Carrasquilla2017Machine}, 
we make use of a fully connected FNN,
which comprises three layers (input, hidden and output) of nodes.
 The number of neurons in the input layer is equal to the number of spins $N=L^d$, where $L$ is the linear size of the lattice and $d$ is the dimensionality. 
The hidden layer is composed of 200 neurons activated by sigmoid functions,
and 
 the output layer has 2 neurons activated by a softmax function.  During training, a cross-entropy cost function is minimized by means of a stochastic gradient descent method with an Adam optimizer~\cite{kingma2015adam}.  
 In order to avoid overfitting, we adopt a dropout regularization ~\cite{srivastava2014dropout},
 and an early stopping strategy.

The input data are spin configurations generated by the  Wolff  algorithm ~\cite{newman1999monte}.
At a given magnetic field $H$ and a given temperature $T$, we prepare $N_{\rm s} \approx 10,000$ samples, and use  80$\%$ of them  for training, 10$\%$ for validation and 10$\%$ for  prediction. 
Random shuffling is applied two to four times to make sure that there are sufficient 
samples at each combination of $(H,T)$ during  training. 

\subsection{Learning second-order phase transitions in two and three dimensions}

\begin{figure}[ht]
\centerline{\includegraphics[width=0.8\columnwidth]{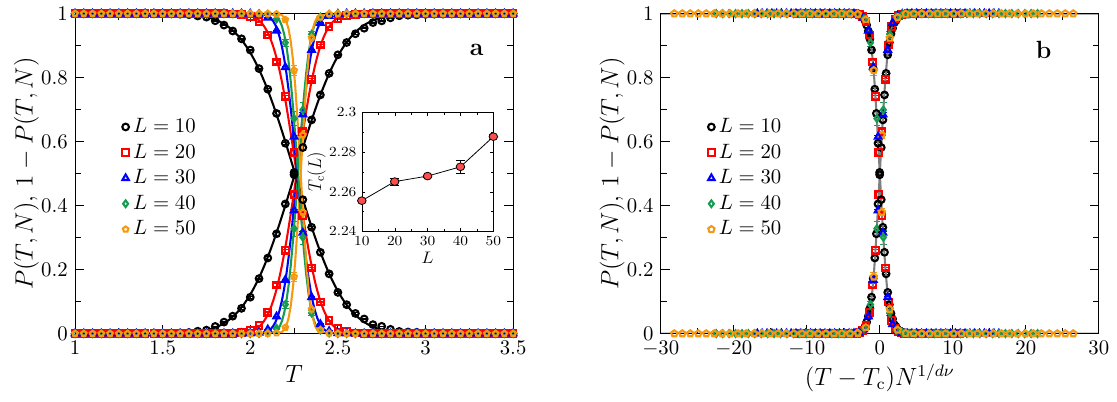}} 
\caption{{
Machine learning the second-order phase transition in the 2D Ising model.} 
Output from machine leaning, $P(T, N)$  and $1-P(T, N)$, as functions of (a) $T$ and (b) $(T-T_{\rm c})N^{1/d\nu}$, for a few different $N=L^2$ at $H=0$. 
The data points are fitted to Eq.~(\ref{eq:error_T}) (lines), where the fitting parameters $T_{\rm c}$ and $w$ are presented in the inset of (a) and Fig.~(\ref{fig:finite_size_Ising}).
The exponent $b=1/d\nu=0.517(4)$ is obtained from fitting $w(N) = w_0 N^{-b}$, and is used in (b).
Error bars represent the standard error of the mean in all figures.
}
\label{fig:Ising_2D_Second}
\end{figure}

\begin{figure}[ht]
\centerline{\includegraphics[width=0.8\columnwidth]{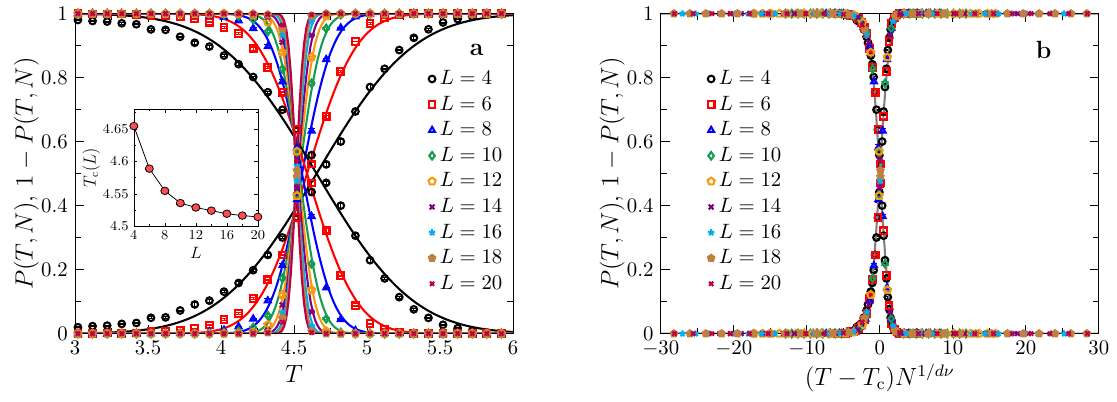}} 
\caption{
{
Machine learning the second-order phase transition in the 3D Ising model.} Output from machine leaning, $P(T, N)$  and $1-P(T, N)$, as functions of (a) $T$ and (b) $(T-T_{\rm c})N^{1/d\nu}$, for a few different $N=L^3$ at $H=0$, where $b=1/d\nu=0.528(4)$. The inset of (a) shows $T_{\rm c}(L)$.
}
\label{fig:Ising_3D_Second}
\end{figure}

In two and three dimensions, a second order phase transition occurs at $T_{\rm c}$ when the temperature is varied under the zero-field condition $H=0$. Previous studies have established the values of $T_{\rm c}$ and $\nu$ (the critical exponent for the divergence of 
correlation length): $T_{\rm c} = 2.26918531421…$~\cite{onsager1944crystal} 
and $\nu=1$~\cite{fisher1967theory} in 2D; $T_{\rm c}\approx4.511528(6)$ 
~\cite{talapov1996magnetization} 
and $\nu\approx0.63012(16)$~\cite{campostrini200225th} in 3D. Supervised machine learning techniques have been well utilized to learn the continuous phase transition in the Ising model  in both 2D~\cite{Carrasquilla2017Machine} and 3D~\cite{zhang2019few}. Here we reproduce these results using our algorithm. 
For this purpose, we generate zero-field ($H=0$) input configurations around $T_{\rm c}$ at  $N_{T}$ different temperature points.
The data points $P(T, N)$ obtained from  machine learning are fitted to  
\beq
P(T, N) = \frac{1}{2} + \frac{1}{2}\erf \left\{\left[T - T_{\rm c}(N)\right]/w(N) \right\},
\label{eq:error_T}
\eeq
where $N = L^d$, $\erf(x)$ is the error function, and  $T_{\rm c}(N)$  and $w(N)$ are two fitting parameters representing the critical temperature and the width of transition region (see Fig.~\ref{fig:Ising_2D_Second} for 2D and Fig.~\ref{fig:Ising_3D_Second} for 3D).  The estimated critical temperatures agree with existing values (see insets of Figs.~\ref{fig:Ising_2D_Second} and~\ref{fig:Ising_3D_Second}).
Next, we examine the  finite-size scaling function Eq.~(2). The scaling function suggests that,
\beq
w(N) \sim N^{-b},
\label{eq:w_scaling}
\eeq
which is used to determine the critical exponent,  $\nu = 0.967(4)$ (or $b = 1/d\nu = 0.517(4)$) in 2D and  $\nu = 0.631(3)$ (or $b  = 0.528(4)$) in 3D (see Fig.~\ref{fig:finite_size_Ising}). These estimations are in a good agreement with the standard values,  $\nu=1$ in 2D~\cite{fisher1967theory} and $\nu\approx0.63012$ in 3D~\cite{campostrini200225th}. Indeed, the data points of $P(T, L)$ for different sizes can be collapsed onto a single curve if they are plotted as a function of the rescaled quantity $(T-T_{\rm c})N^{1/d\nu}$ (see Figs.~\ref{fig:Ising_2D_Second}(b) and~\ref{fig:Ising_3D_Second}(b)).

\subsection{Learning first-order phase transitions in two and three dimensions}

\begin{figure}[ht]
\centerline{\includegraphics[width=0.8\columnwidth]{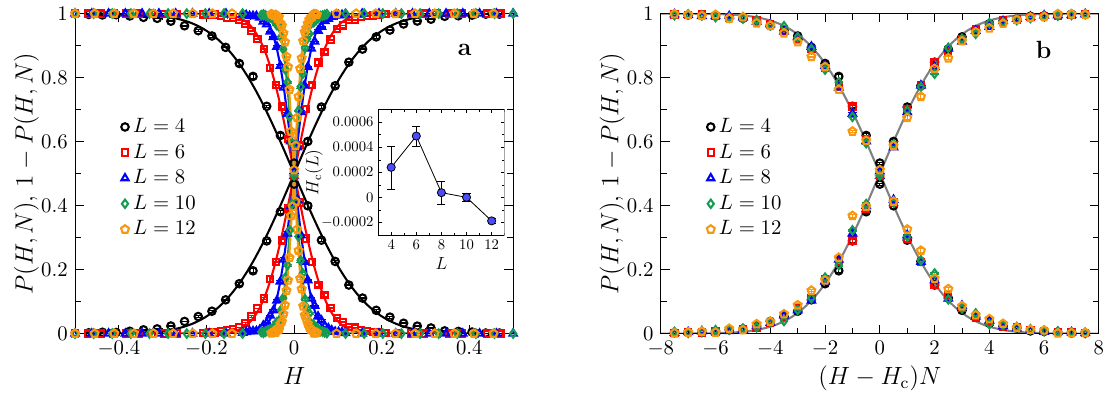}} 
\caption{
{Machine learning the first-order phase transition in the 2D Ising model.}
Output from machine leaning, $P(H, N)$  and $1-P(H, N)$, as functions of (a) $H$ and (b) $(H-H_{\rm c})N$, for a few different $N=L^2$. The inset of (a) shows $H_{\rm c}(L)$.
}
\label{fig:Ising_2D_first}
\end{figure}

\begin{figure}[ht]
\centerline{\includegraphics[width=0.8\columnwidth]{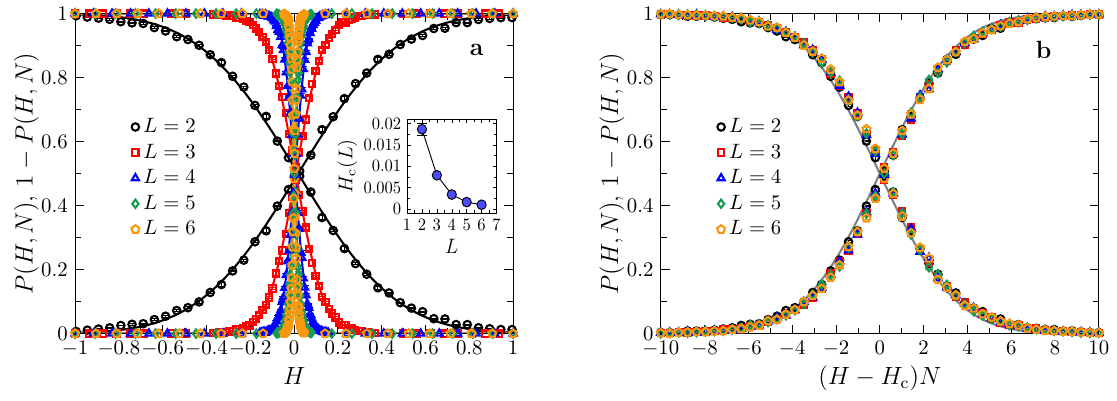}} 
\caption{
{Machine learning the first-order phase transition in the 3D Ising model.}
Output from machine leaning, $P(H, N)$  and $1-P(H, N)$, as functions of (a) $H$ and (b) $(H-H_{\rm c})N$, for a few different $N=L^3$. The inset of (a) shows $H_{\rm c}(L)$.}
\label{fig:Ising_3D_first}
\end{figure}

For a fixed temperature $T<T_{\rm c}$, a first-order phase transition occurs at $H_{\rm c}=0$ when $H$ is varied. 
To our knowledge, the finite-size scaling Eq.~(2) of the first-order phase transition in the Ising model has not been systematically studied yet within the machine learning framework.  In this study, we show that the expected scaling Eq.~(2) is fully consistent with our data. We
set $k_{\rm B} T/J=2.1$ for the 2D model and $k_{\rm B} T/J=4.0$ for the 3D model, where $J$ is the interaction constant. Configurations are generated at $N_H$ different external fields around $H_{\rm c}=0$, with positive and negative  fields evenly divided.
The data points $P(H, N)$ obtained from  machine learning are fitted to  
\beq
P(H, N) = \frac{1}{2} + \frac{1}{2}\erf \left\{\left[H - H_{\rm c}(N)\right]/w(N) \right\}.
\label{eq:error_H}
\eeq
As shown in Figs.~\ref{fig:Ising_2D_first} and ~\ref{fig:Ising_3D_first}, the estimated transition field is close to  $H_{\rm c} = 0$. Furthermore, we obtain $b \approx 0.96(2)$ for the 2D model, and $b \approx 1.012(4)$ for the 3D model, which are  consistent with the expected value $b=1$~\cite{binder1987theory}. 

\subsection{Distinguishing between first- and second-order phase transitions}

\begin{figure*}[ht]
\centerline{\includegraphics[width=0.7\columnwidth]{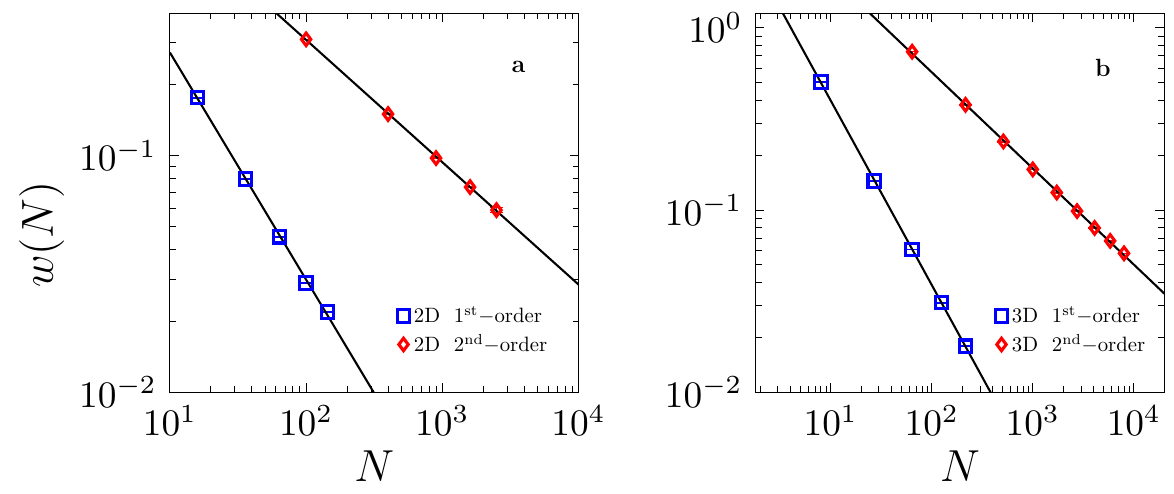}} 
\caption{{Comparing the finite size scalings of first- and second-order phase transitions in the Ising model.}
The data of transition width $w(N)$ are obtained from Figs.~(\ref{fig:Ising_2D_Second}-\ref{fig:Ising_3D_first}).
The exponent $b$ is obtained from fitting $w(N) = w_0 N^{-b}$ (lines).  (a) In the 2D Ising model, we obtain $b = 0.96(2)$ for the first-order phase transition, and $b = 1/d\nu = 0.517(4)$ (i.e., $\nu = 0.967(4)$) for the second-order phase transition. (b) In the 3D Ising model, we obtain $b = 1.012(4)$ for the first-order phase transition, and $b  = 0.528(4)$ (i.e., $\nu = 0.631(3)$) for the second-order phase transition.}
\label{fig:finite_size_Ising}
\end{figure*}

Based on above analyses, we confirm that, by utilizing the scaling function Eq.~(2), the original machine learning approach proposed in~\cite{Carrasquilla2017Machine} can be generalized to identify both first- and second-order phase transitions,  in the standard Ising model. Very importantly, the order of phase transition can be identified because the finite-size exponents $b$ in Eq.~(2) are distinguishable within the numerical accuracy for first- and second-order phase transitions. As shown in  Fig.~\ref{fig:finite_size_Ising}, $b=1$ for first-order phase transitions (without considering the effect of disorder), and $b=1/d\nu$ for second-order phase transitions. While the phase transitions in the Ising model are in equilibrium, we show that the approach can be further generalized to non-equilibrium  first-order (melting transition) and second-order (Gardner transition) phase transitions in disordered systems such as glasses (see {Fig.~5}).

\section{Additional results for the melting transition}
\label{sec:additional_melting}


\subsection{Dependence of the melting temperature on the decompression rate}
\label{sec:melting_rate}

\begin{figure}[ht]
\centerline{\includegraphics[width=0.35\columnwidth]{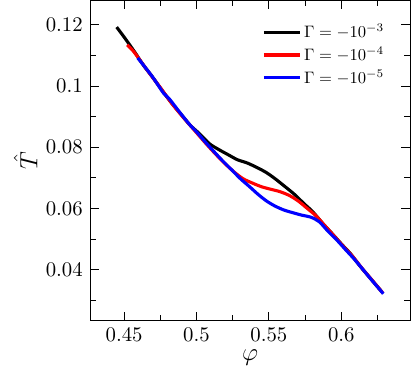}} 
\caption{{Evolution of temperature $\hat{T}$ as a function of volume fraction $\varphi$ under decompression, for a few different decompression rate $\Gamma$}.
The system is composed of  $N=125$ particles, and is decompressed from ultra-stable states at $\{\varphi_{\rm g} = 0.63, \hat{T}_{\rm g} = 0.033 \}$. The liquid and glass EOSs are connected around the melting temperature $\hat{T}_{\rm m}$, which decreases with slower decompression. 
}
\label{fig:rate_dependence}
\end{figure}

The $\hat{T}-\varphi$ EOSs of ultra-stable glasses in Fig.~\ref{fig:rate_dependence} show that  
the melting transition temperature $\hat{T}_{\rm m}$ decreases with slower decompression. 
It is expected that, in the limit $\Gamma \to 0$, 
the hysteresis in EOS will disappear and the glass melting will become a continuous crossover. 
However, for the range of $\Gamma$ relevant to this study,  the discontinuous  feature remains.
In the main text, we examine the finite-size effect for a fixed decompression rate $\Gamma = -10^{-4}$, and do not further discuss the rate-dependence. 

\subsection{Independence of learning results on the blanking window} 
\label{sec:blanking_melting}

\begin{figure}[ht]
\centerline{\includegraphics[width=0.8\columnwidth]{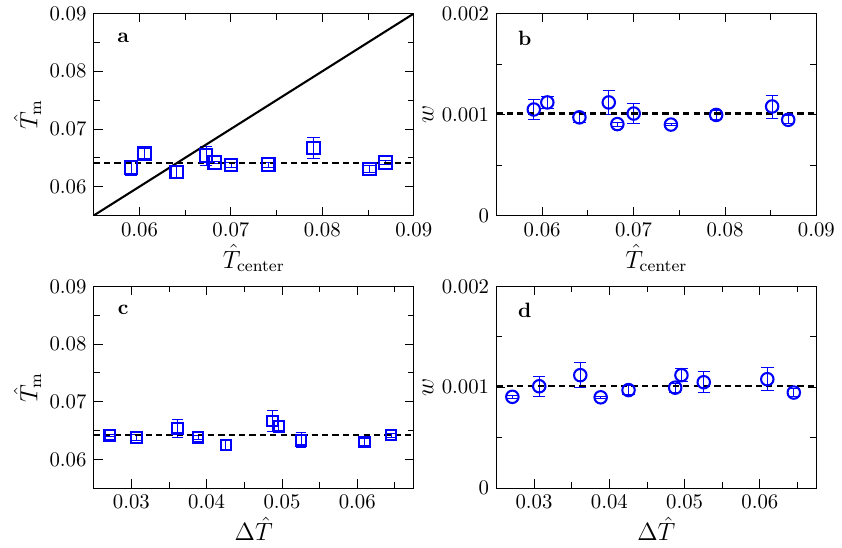}} 
\caption{ Independence of $\hat{T}_{\rm m}$ and $w$ on the blanking window $[\hat{T}_2,\hat{T}_1]$, for the melting transition. The two-level NNN is trained using a few different combinations of $\hat{T}_1$ and $\hat{T}_2$, for  $N=2000$ and $\Gamma = 10^{-4}$. The predicted $\hat{T}_{\rm m}$ and $w$ are plotted as functions of $\hat{T}_{\text{center}}$ and $\Delta \hat{T}$. The  horizontal dashed lines are $\hat{T}_{\rm m}(N=2000)=0.064$ and $w(N=2000)=0.001$ obtained for $T_1 = 0.083$ and $T_2 = 0.053$. 
The solid line in (a) represents $\hat{T}_{\rm m} = \hat{T}_{\rm center}$.
The same data in (a) are plotted in {Fig.~9(a)}.
}
\label{fig:melting_blanking_window}
\end{figure}

During training, the samples at $\hat{T} > \hat{T}_1$ and  $\hat{T} < \hat{T}_2$ are labeled as in the liquid and glass phases respectively.
The samples in the blanking window $[\hat{T}_2,\hat{T}_1]$ are not used. Figure~\ref{fig:melting_blanking_window} shows that the machine predicted melting temperature $\hat{T}_{\rm m}$  and transition width $w$ are independent of the  blanking window (more specifically, the center of window $\hat{T}_{\rm center} = (\hat{T}_1 +\hat{T}_2)/2$ and the width of window $\Delta \hat{T} = \hat{T}_1 - \hat{T}_2$). Note that, obviously we should require $\hat{T}_{\rm m}$ to be inside of the blanking window, i.e., $\hat{T}_2 < \hat{T}_{\rm m} < \hat{T}_1$. With this restriction, the choice of blanking window is flexible. 

\section{Additional results for the Gardner transition}
\label{sec:additional_Gardner}



\subsection{Choice of input data} 
\label{sec:pre_Gardner}

\begin{figure}[ht]
\centerline{\includegraphics[width=0.4\columnwidth]{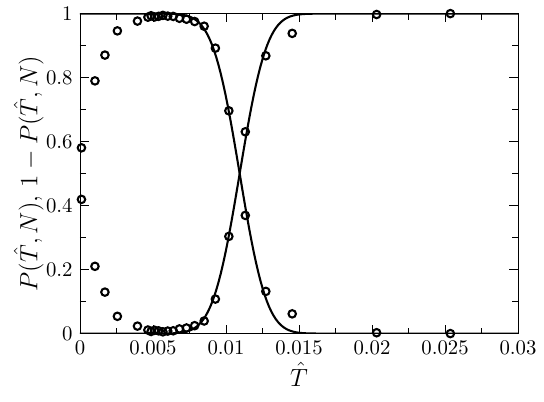}} 
\caption{Failure to learn the Gardner transition using 
$\{ |\vec{r}_i^{AB}|^2 \}$ as input data.
The test is performed 
for $N=2000$ systems. 
}
\label{fig:Gardner_DeltaAB}
\end{figure}

According to the predictions from the mean-field glass theory~\cite{parisi2020theory, charbonneau2017glass}, the features of stable and marginally stable phases  are encoded in $\{\Delta_i \}$, and thus in principle one should be able to use $\{ |\vec{r}_i^{AB}|^2 \}$ as input data to train networks. However, in practice, the network fails to correctly identify both phases, when 
 $\{ |\vec{r}_i^{AB}|^2 \}$
are used as input. We find that (see Fig.~\ref{fig:Gardner_DeltaAB}),  as $\hat{T} \to 0$, the predicted probability  $P \approx 0.5$, 
while physically we expect $P \approx 1$ (the samples should all belong to the marginally stable phase at sufficiently low $\hat{T}$). On the other hand, correct and robust predictions are obtained when $\{u^{AB}_i = \frac{|\vec{r}_i^{AB}|^2}{\Delta_i}  - 1 \}$ 
are used as input.
Indeed,  the cage size $\Delta_i = \langle |\vec{r}_i^{AB}|^2 \rangle_{
\rm r} $ generally becomes smaller with decreasing $\hat{T}$,
but this effect is independent of the physics of Gardner transition.
The purpose of normalization is to remove this effect. 

\subsection{Independence of learning results on the blanking window} 
\label{sec:blanking_Gardner}

\begin{figure}[ht]
\centerline{\includegraphics[width=0.8\columnwidth]{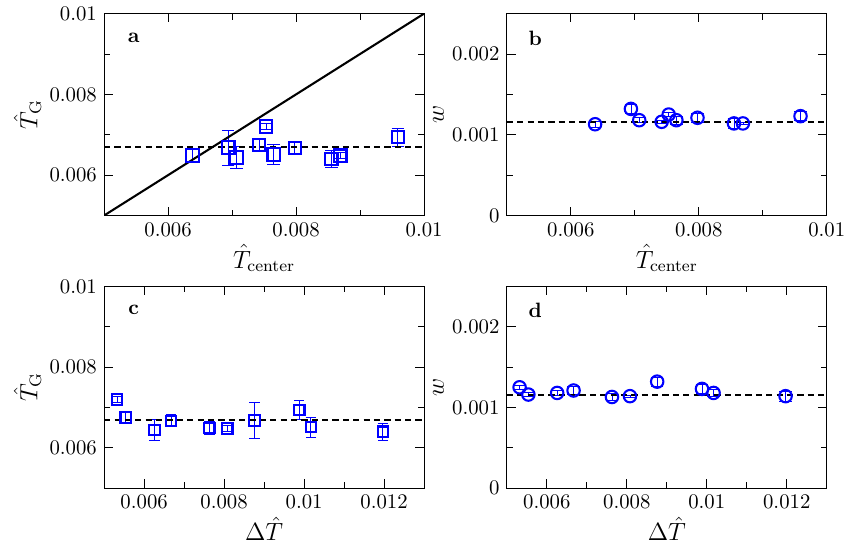}} 
\caption{Independence of $\hat{T}_{\rm G}$ and $w$ on the blanking window $[\hat{T}_2,\hat{T}_1]$, for the Gardner transition. 
The predicted $\hat{T_{\rm G}}$ and $w$ are plotted as function of $\hat{T}_{\text{center}}$ and $\Delta \hat{T}$ (for $N=8000$ systems). 
The horizontal dashed lines are $\hat{T}_{\rm G}(N=8000)=0.0067$ and $w(N=8000)=0.0012$ obtained for $T_1 = 0.011$ and $T_2 = 0.0045$. 
The solid line in (a) represents $\hat{T}_{\rm G} = \hat{T}_{\rm center}$.
The same data in (a) are plotted in {Fig.~9(b)}.
}
\label{fig:gardner_blanking_window}
\end{figure}

Figure~\ref{fig:gardner_blanking_window} shows that the machine predicted Gardner transition temperature $\hat{T}_{\rm G}$ and the transition width $w$ are independent of the choice of blanking window. 

\section{Additional dynamical data for the glass transition}
\label{sec:additional_glass}

\begin{figure}[ht]
\centerline{\includegraphics[width=0.7\columnwidth]{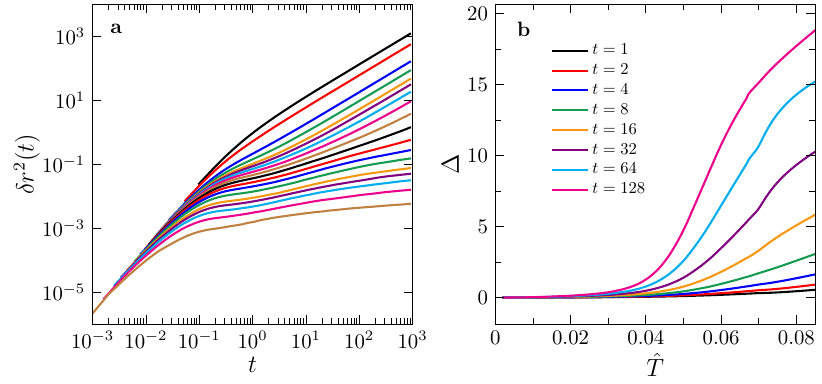}}
\caption{{Dynamical data of the glass transition.}
(a) MSD data at different densities ($N=500$): from top to bottom, $\varphi$ = 0.3, 0.4, 0.5, 0.53, 0.55, 0.56, 0.57, 0.58, 0.59, 0.6, 0.61, 0.62, 0.63, 0.64, 0.645, 0.65, 0.655, 0.66 ($\hat{T}=$ 0.27, 0.16, 0.084, 0.069, 0.059, 0.054, 0.050, 0.045, 0.040, 0.036, 0.030, 0.025, 0.019, 0.014, 0.011, 0.0078, 0.0049, 0.0018). 
(b) Average cage size $\Delta$ as a function of 
$\hat{T}$
for a few different measurement time $t$.
}
\label{fig:dynamics_glass}
\end{figure}

In Fig.~\ref{fig:dynamics_glass}, we plot the data of mean-squared displacement (MSD), 
\beq
\delta r^2 (t) = \frac{1}{N} \overline{\sum_{i=1}^N |\vec{r}_i(t) - \vec{r}_i(0)|^2},
\eeq
 around the liquid to ordinary glass transition,
where $\vec{r}_i(0)$ is the position of particle $i$ right after compression with a rate $\Gamma = 10^{-3}$, and $\vec{r}_i(t)$ is the position at  time $t$ (we set $t=0$ and $\Gamma = 0$ after compression). The dynamics clearly slow down with decreasing $\hat{T}$, but activated processes are non-negligible since the MSD is not completely flat even at low  $\hat{T}$.
Figure~\ref{fig:dynamics_glass}(b) shows that the average cage size $\Delta$   changes smoothly with 
$\hat{T}$
and depends sensitively on the measurement time $t$. 
For a comparison, see the MSD of ultra-stable glasses in Fig.~2 of Ref.~\cite{berthier2016growing}.


\section{Learning a toy model}

We artificially construct a toy model consisting of two phases, which are represented by two distribution functions respectively: a single Gaussian distribution $p_{\rm I}(x) = p_{\rm G}(x; \mu, \sigma)$ for phase I, and  a two-Gaussian distribution $p_{\rm II}(x) = \frac{1}{2} p_{\rm G}(x; \mu_{1}, \sigma_{1}) + \frac{1}{2} p_{\rm G}(x; \mu_{2}, \sigma_{2})$ for phase II, 
where $p_{\rm G}(x; \mu, \sigma) = \frac{1}{\sigma \sqrt{2 \pi}} e^{-\frac{(x- \mu)^2}{2\sigma^2}}$ is the standard Gaussian (normal) distribution of mean $\mu$ and  variance $\sigma^2$. We choose the parameters such that the means and variances are identical for the two distributions, but the skewnesses $\langle \left( \frac{x-\mu}{\sigma}\right) ^3 \rangle$ are different (see Table~\ref{tab:toy}): { $\mu = 0, \sigma = 1.768$ in $p_{\rm I}(x)$,  and $\mu_1 = -1, \sigma_1 = 2, \mu_2 = 1, \sigma_2 = 0.5$ in $p_{\rm II}(x)$.}

The setup for the machine learning algorithm is similar to that illustrated in Fig.~1. We use $N=1$ ``particle", whose feature is described by  an input vector $\vec{V}$ of $M_{\rm r} = 1000$  random numbers drawn from  $p_{\rm I}(x)$ or $p_{\rm II}(x)$. Thus $\vec{V}$ is a representation of the probability distribution function for each phase. The front-connected small network maps the distribution $p_{\rm I}(x)$ or $p_{\rm II}(x)$ to a single scalar ``order parameter" $q$. The two-level NNN is trained by {$N_{\rm s}^{\rm train} = 15000$} samples, and makes the phase prediction for {$N_{\rm s}^{\rm pred} = 2500$} samples. The rest of the method is equivalent to that for learning the glass model. 

The Pearson correlation coefficients $r$ between the first three moments and the parameter $q$ are computed, taking account of both phases (see Table~\ref{tab:toy}). The predicted $q$  is strongly correlated to the skewness  ($r \approx 1$), while its correlation  to the mean or the variance is negligible ($r \approx 0$). This exercise shows that our  method can correctly extract higher-order statistical moments from the  input data  when the mean and variance are trivial.

\begin{table}[h]
\caption{{
Machine learning results for the toy model. Presented are the first three moments (mean, variance and skewness) of probability distribution functions $p_{\rm I}(x)$ and $p_{\rm II}(x)$, and the  Pearson correlation coefficient $r$ between the moment and the parameter $q$ learned by the small network.}}
\begin{tabular}{ c c c c }
\hline
\hline
&  $p_{\rm I}(x)$ & $p_{\rm II}(x)$ &   $r$  \\
\hline
mean   & 0 & 0   & -0.018   \\
variance    & 3.125 & 3.125   & 0.109 \\
skewness   & 0           & -5.625    & 0.957 \\
\hline
\hline  
\end{tabular}
\label{tab:toy}
\end{table}

 \color{black}

\clearpage

\end{document}